\documentclass[%
 aip,
 amsmath,amssymb,
 reprint,%
]{revtex4-1}

\usepackage{graphicx}
\usepackage{dcolumn}
\usepackage{bm}

\usepackage[utf8]{inputenc}
\usepackage[T1]{fontenc}
\usepackage{mathptmx}
\usepackage{etoolbox}

\makeatletter
\def\@email#1#2{%
 \endgroup
 \patchcmd{\titleblock@produce}
  {\frontmatter@RRAPformat}
  {\frontmatter@RRAPformat{\produce@RRAP{*#1\href{mailto:#2}{#2}}}\frontmatter@RRAPformat}
  {}{}
}%
\makeatother
\begin{document}

\preprint{AIP/123-QED}

\title[Spectroscopic measurements of graphite electrode erosion on the ZaP-HD
sheared-flow-stabilized Z-pinch device
]{Spectroscopic measurements of graphite electrode erosion on the ZaP-HD
sheared-flow-stabilized Z-pinch device
}
\author{A. A. Khairi}
 \email{aqilk@uw.edu.}

\author{E. Lian}%
 \author{U. Shumlak}
\affiliation{ 
Aerospace and Energetics Research Program, University of Washington, Seattle, WA 98195-2400
}%


\date{\today}

\begin{abstract}
The ionizations per photon, or S/XB, method uses spectroscopic measurements of radiating impurity ions to determine the influx from a solid surface. It is highly useful as a non-perturbing, in-situ measure of the gross erosion flux of plasma-facing components (PFCs). In sheared-flow-stabilized (SFS) Z-pinch devices, the electrode supplies the plasma current and directly contacts the core Z-pinch plasma. Electrode erosion due to the large particle and heat fluxes affects electrode durability, which is an important factor in existing and future devices. An improved understanding of these plasma-electrode interactions is required, in particular as energy density increases. Experiments on the ZaP-HD device investigate erosion of the graphite electrode by applying the S/XB method for C-III emission at 229.7~nm. The S/XB coefficients are determined from electron density and temperature profiles obtained from Digital Holographic Interferometry (DHI) measurements. An approach for expanding these profiles to represent plasma contacting the electrode is described. In both cases, the measured erosion fluxes are on the order of 10$^{30}$-10$^{31}$~atoms~m$^{-2}$s$^{-1}$. These values are significantly larger than the expected erosion flux due to physical sputtering of H$^+$ ions on carbon, but are comparable to theoretical sublimation fluxes. This suggests that the source of carbon erosion flux is primarily from sublimation as opposed to sputtering. The dominance of sublimation over sputtering processes implies a difference in energy of the eroded neutrals which may provide insight on redeposition and net erosion behavior. 
\end{abstract}

\maketitle


\section{\label{sec:intro}Introduction}
All fusion devices contend with the challenge of plasma-material interactions (PMI) in some form, one characterized by a diverse and complex set of physical processes. These interactions affect the performance of fusion devices, causing lifetime-limiting erosion damage and introducing impurities that reduce the plasma temperature. The importance of PMI in fusion devices is underscored by extensive research over several decades, \cite{Zinkle2013, Linsmeier_2017, Linke_2019_MRE} although with a large focus on toroidal configurations such as the tokamak and stellarator. The understanding of PMI effects specific to toroidal devices is therefore more mature compared to alternative approaches such as the sheared-flow-stabilized (SFS) Z pinch, where a solid electrode directly contacts the core plasma. The electrode is exposed to high particle and heat fluxes and must sustain a large plasma current density. The resulting plasma-electrode interactions will have important ramifications as the concept scales to power plant conditions, and work remains to sufficiently investigate their effects at high plasma temperature and current densities. \cite{Shumlak2020} 

The SFS Z-pinch concept faces unique challenges with regard to PMI. In the existing literature, a strong focus is placed on the low temperature regime ($\sim$10~eV) expected at the ITER divertor. \cite{Ikeda_2007} Experiments on toroidal devices and dedicated linear testbeds \cite{Kreter2011} serve to replicate this and other ITER-like edge plasma conditions. \cite{Linsmeier_2017_Facilities} By contrast, SFS Z-pinch electron temperatures at the electrode can be up to two orders of magnitude greater. \cite{Shumlak2017} Although the SFS Z pinch does operate in a higher density regime, existing PMI testbeds are capable of long pulse durations or steady-state operation to provide large particle fluences. Nevertheless, the SFS Z pinch can be further distinguished by several aspects. Plasma exposure is concentrated over a relatively small contact area on the electrode when compared to a divertor or interior first wall. In addition to the heat flux from the plasma, electrode heating also arises from driving large currents. Ion impact energies are enhanced by the applied electrode voltage, which increases the sheath potential drop. These factors lead to intense plasma-electrode interactions that result in material erosion. Erosion damage impacts electrode durability and overall device reliability, and has been identified as a key concern for a future SFS Z-pinch power plant. \cite{Thompson_2023_FST, Thompson_2023_POP} Similar erosion concerns are relevant to arc discharges, \cite{Kimblin_1974, Brown_1990} although for much lower plasma temperatures. This motivates efforts to characterize the erosion behavior in the high temperature, high current density environment of SFS Z-pinch devices.

This paper describes the application of the S/XB method for inferring the eroded carbon flux on the graphite electrode of the ZaP-HD SFS Z-pinch device. This represents novel measurements of erosion on a solid surface exposed to a high temperature plasma and large current densities. An approach is presented for determining the appropriate S/XB coefficients based on measured radial profiles of the electron number density and temperature. These inferred erosion fluxes are then compared to theoretical values of the erosion flux to evaluate the potential underlying physical mechanisms. Finally, the measured density profiles are expanded to better represent plasma profiles on the electrode surface where the S/XB measurement is taken.

\subsection{\label{sec:introsxb}Measuring erosion with S/XB spectroscopy}

Spectroscopy has been a valuable tool for monitoring the erosion of solid components in fusion devices, often by measuring the absolute intensities of vacuum ultraviolet (VUV) and visible wavelength emissions. Early implementation occurred in the 1980s on various tokamaks \cite{DeMichelis_1984, Morgan1985, Behringer_1986RSI, HINTZ1984229, FUSSMANN1984350} and continues in more recent experiments. \cite{RUDAKOV2017196, Abrams_2021, James_2013} The spectroscopic technique applied is known as the ionizations per photon, or S/XB, method. In these examples, the absolute emission intensity of eroded ions was measured and converted to a photon flux. The photon flux depends on the excitation rate of the species under investigation. The population of that species is determined by electron impact ionization of neutrals removed from the surface of the plasma-facing component (PFC). Therefore, the flux of the eroded neutrals can be inferred by applying the appropriate ionization and excitation rates that would result in the measured photon flux. These rates depend on the cross-section for excitation and ionization for the species of investigation, which ultimately depend on the local plasma temperature and density.

The S/XB method has the advantage of providing an in-situ erosion measurement. The erosion behavior can therefore be characterized during operation. By contrast, ex-situ erosion measurements involve the removal of sample material after plasma exposure, which can be a time-consuming, complex, and disruptive process. This can be addressed by the use of dedicated experimental testbeds or custom apparatus such as DiMES \cite{WONG2007276} on the DIII-D tokamak. However, ex-situ measurements are integrated over numerous plasma pulses, which prevents the assessment of particular input settings. This motivates the use of the S/XB method to provide an erosion measurement on a per-pulse basis. In practice, a combination of in-situ and ex-situ techniques is preferred, which would enable comparison of campaign-averaged effects with individual pulses. It is worth noting that the S/XB method is limited to a gross measurement of the eroded flux, since the ionized neutrals may return to the surface through redeposition. A net erosion measurement using the S/XB method requires knowledge of the redeposition behavior, which can be represented by a redeposition factor. This factor may be calculated \cite{Naujoks_1996} or determined empirically. \cite{ROTH1995231} The redeposition behavior on ZaP-HD requires further investigation, therefore the results presented in this paper are limited to values of the gross eroded flux. 

\subsection{\label{sec:physics}Physics principles}

\begin{figure}[h]
\includegraphics[scale=0.58]{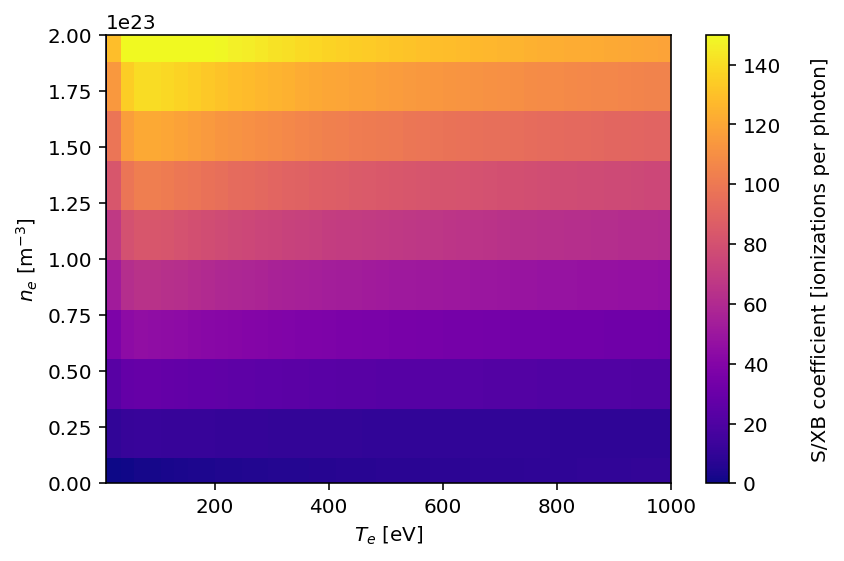}
\caption{\label{fig:sxbcoeffs}S/XB coefficients obtained from the ADAS database for the 229.7 nm emission of the C-III ion, showing variation for the range of typical ZaP-HD plasma parameters. These coefficients are used to convert the measured photon flux of C-III emission into an inferred erosion flux of carbon neutrals.}
\label{sxbplot}
\end{figure}

The S/XB method assumes that excitation and ionization occur solely through electron impact. For electron energies comparable to the excitation and ionization potentials (several~eV), the cross-sections for inelastic electron collisions are several orders of magnitude larger than that for heavy particle collisions such as charge exchange. In the following expressions, $A$ represents a neutral atom in the ground state, $A^*$  represents an excited state, and $A^+$ represents an ion. Electron impact excitation of bound electrons is described by
\begin{equation}
A + e \rightarrow A^* + e,
\end{equation}
while electron impact ionization is described by
\begin{equation}
A + e \rightarrow A^+ + e + e.
\end{equation}
The de-excitation process is restricted to spontaneous emission of a photon with frequency $\nu$:
\begin{equation}
A^* \rightarrow A + h\nu.
\end{equation}
This frequency often corresponds to wavelengths in the visible or UV range, which is readily accessed by spectroscopy. If the ion of study penetrates further into the plasma, it will undergo further ionization. As a result, lower ionization states tend to be more localized to their source and therefore more representative of the erosion flux. However, depending on the local plasma parameters, lower ionization states may be less accessible for measurement due to their low ionization energies. The rates associated with these processes are combined into a coefficient composed of the ratio $S/XB$, where $S$ is the ionization rate, $X$ is the excitation rate, and $B$ is the branching ratio. The S/XB coefficient converts a photon flux into an eroded particle flux by the expression
\begin{equation}
    \Gamma = 4\pi\left( \frac{S}{XB} \right)I,
    \label{sxbgen}
\end{equation}
where $\Gamma$ is the eroded atom flux (atoms~m$^{-2}$s$^{-1}$), and $I$ is the line-integrated photon flux of the target ion (photons~m$^{-2}$s$^{-1}$sr$^{-1}$). Collisional-radiative models have been developed \cite{Behringer_1989} to obtain values of the S/XB coefficient. For this study, the emission of C-III at 229.7~nm was selected for its strong intensity and isolation from other line radiation. This wavelength has been used previously for spectroscopic measurements on the ZaP \cite{DenHartog_2001_RSI} and ZaP-HD \cite{Forbes_2020_RSI} SFS Z-pinch devices. S/XB coefficients for C-III were obtained from the Atomic Data and Analysis Structure (ADAS) database. \cite{Summers_2011_ADAS} These are shown in Fig.~\ref{sxbplot} for plasma parameters observed on ZaP-HD. Due to this variation, application of the S/XB method depends on accurate knowledge of the local plasma temperature and density.
\section{\label{sec:expsetup}Experimental setup on ZaP-HD}

\begin{figure*}
\includegraphics[scale=0.85]{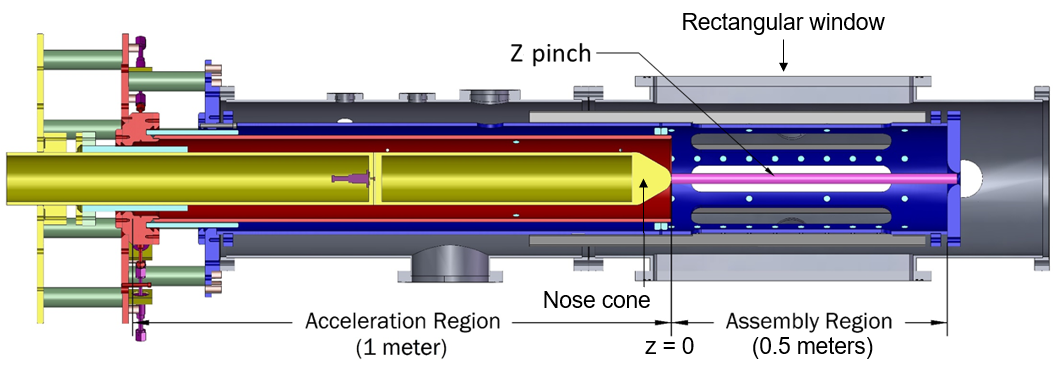}
\caption{\label{fig:zaphd}Cross-sectional machine drawing of the ZaP-HD SFS Z-pinch device showing the Acceleration Region where the plasma is generated and accelerated and the Assembly Region where the Z-pinch plasma is formed and compressed. The Z-pinch plasma (shown schematically in magenta) directly contacts the electrode nose cone. Four rectangular fused silica windows in the Assembly Region provide optical access to the Z-pinch plasma and the electrode for erosion measurements.}
\end{figure*}

\subsection{\label{sec:zaphdsec}The ZaP-HD Experiment}

The ZaP-HD experiment at the University of Washington (UW) investigates sheared-flow stabilization of Z-pinch plasmas at high temperature and density. This follows strong evidence for sheared-flow-stabilization observed on the ZaP experiment, \cite{Shumlak_2001_PRL, Golingo_2005, Shumlak_2009} which supports theoretical predictions. \cite{Shumlak1995} Further development of the concept for commercial fusion applications is underway. \cite{Levitt_2023}

On ZaP-HD, three coaxial electrodes define an Acceleration Region and an Assembly Region, illustrated in Fig.~\ref{fig:zaphd}. In the Acceleration Region, neutral gas is puffed into the annular volume between the middle and inner electrodes. The first of two capacitor banks applies a voltage between these electrodes, ionizing the gas and driving a current, inducing a Lorentz force that accelerates the plasma downstream to the Assembly Region. When the plasma reaches the Assembly Region, the second capacitor bank applies a voltage between the outer and inner electrodes, driving additional current that compresses the plasma on axis. The Z-pinch plasma directly contacts the electrode at the nose cone. Optical access is provided by four rectangular fused silica windows. Decoupling the plasma ionization and compression processes provides increased control of the input parameters and enables higher pinch currents to be driven. Previous measurements indicated stability for 50-60~$\mu$s, peak electron temperatures up to 1 keV, peak electron densities up to 10$^{24}$~m$^{-3}$, and pinch radii of 0.3~cm. \cite{Shumlak2017}

An analysis of Z-pinch equilibrium yields scaling relations that describe the variation of temperature, density and pinch radius with the current. This enables calculation of radial profiles that are used to determine the S/XB coefficients, which is described in Sec.~\ref{sec:SXBsel}. Z-pinch equilibrium is described by a radial force balance between the plasma pressure and the self-generated azimuthal magnetic field,
\begin{equation}
    \frac{d}{dr}\left (n_ik_BT_i + n_ek_BT_e \right) = -\frac{B_\theta}{\mu_0r}\frac{d}{dr}\left( rB_\theta \right),
    \label{radeq}
\end{equation}
where $n_i$ and $n_e$ are the ion and electron number densities, $T_i$ and $T_e$ are the ion and electron temperatures, and $B_\theta$ is the azimuthal magnetic field. Integration of Eq.~(\ref{radeq}) over the plasma volume results in the Bennett relation, \cite{Bennett1934}
\begin{equation}
    \left( 1+Z \right)N_ik_B\langle T \rangle = \frac{\mu_0I^2}{8\pi},
\end{equation}
which describes scaling of the plasma density and temperature to the plasma current $I$. Here, $Z$ is the ionization state, $N_i$ is the linear ion number density, and $\langle T \rangle$ is the average temperature assuming thermal equilibrium between ions and electrons. For an adiabatically compressed Z-pinch plasma, and using the "sharp pinch" model \cite{NEWCOMB1960232} for an equilibrium with uniform density, scaling relations \cite{Shumlak2012_FST, Shumlak2017, Shumlak2020} for the temperature, density, and pinch radius are derived as
\begin{equation}
    \frac{T_2}{T_1} = \left( \frac{I_2}{I_1} \right)^2 \frac{N_1}{N_2},
\end{equation}
\begin{equation}
    \frac{n_2}{n_1} = \left( \frac{I_2}{I_1} \right)^\frac{2}{\gamma-1}\left( \frac{N_1}{N_2} \right)^\frac{1}{\gamma-1},
\end{equation}
\begin{equation}
    \frac{a_2}{a_1} = \left( \frac{I_1}{I_2} \right)^\frac{1}{\gamma-1} \left( \frac{N_2}{N_1} \right)^\frac{\gamma}{2(\gamma-1)}.
\end{equation}
In the above expressions, $N$ is the linear number density defined as
\begin{equation}
    N = \int_0^a{2\pi n(r) rdr},
    \label{lineardensityeq}
\end{equation}
for the pinch radius $a$. These relations indicate that increasing the current for a Z pinch with constant linear density increases the temperature and density, while decreasing the plasma radius.

\subsection{\label{sec:sxbsetup}S/XB Spectroscopy Setup}

\begin{figure}[h]
\includegraphics[scale=0.8]{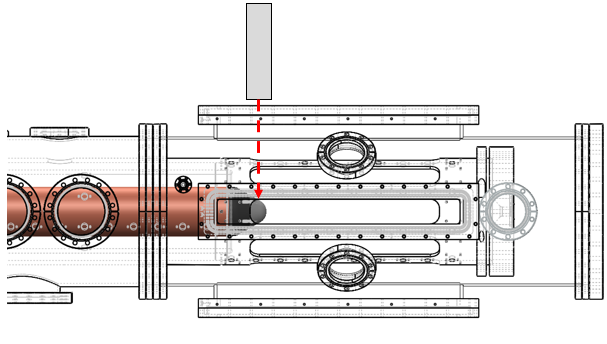}
\caption{\label{fig:telescope_position}Recent modifications to ZaP-HD placed the electrode further downstream, simplifying optical access with a direct line of sight to the telescope used for S/XB spectroscopy.}
\end{figure}

\begin{figure}[h]
\includegraphics[scale=0.6]{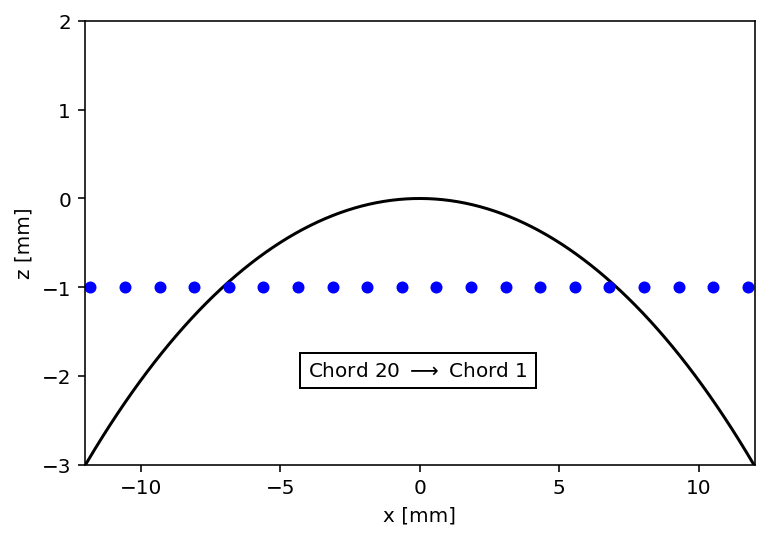}
\caption{\label{fig:chord_position} Impact parameters of the 20 spectroscopy chords on the electrode nose cone profile. Plasma flows downstream in the positive $z$ direction. The central axis of ZaP-HD is along $x = 0$. The measurement chords span a total of 23.6~mm across the nose cone. All but the four outer chords on either side of the array have a line of sight which terminates on the electrode surface.}
\end{figure}

This diagnostic uses a similar setup as previous spectroscopy measurements on ZaP and ZaP-HD. \cite{Golingo_RSI_2003, Forbes_2020_RSI} UV emission from the plasma is transmitted through the windows and focused by a telecentric telescope \cite{DenHartog_2001_RSI} onto a bundle of optical fibers. The light is transmitted to an Acton SpectraPro 500i Czerny-Turner spectrometer with a UV diffraction grating. The fiber bundle is composed of 20 individual multimode fibers with 400~$\mu$m diameters arranged in a linear array. Within the plasma, these chords span a total of 23.6 mm. The detector is a PI-MAX 4 ICCD camera with 1024~x~1024~pixel resolution. Recent modifications on ZaP-HD positioned the tip of the electrode 8~cm downstream of the point labeled as $z=0$ in Fig.~\ref{fig:zaphd}. As shown in Fig.~\ref{fig:telescope_position}, this change was done to facilitate direct lines of sight to the electrode for spectroscopy. Figure~\ref{fig:chord_position} shows the impact parameter $x$ of the 20 chords overlaid on the profile of the electrode nose cone. For chords that terminate on the electrode surface, the chord-integrated intensity includes emission of C-III near the surface as well as emission along the entire line of sight. To isolate the C-III emission measured at the nose cone surface, the intensity of the four outermost chords, which do not terminate on the nose cone, are subtracted as a background measurement.

\section{\label{sec:dataprocessing}Calibration and data processing}

\subsection{\label{sec:ahscap}Absolute Irradiance Calibration}

\begin{figure}[h]
\includegraphics[scale=0.6]{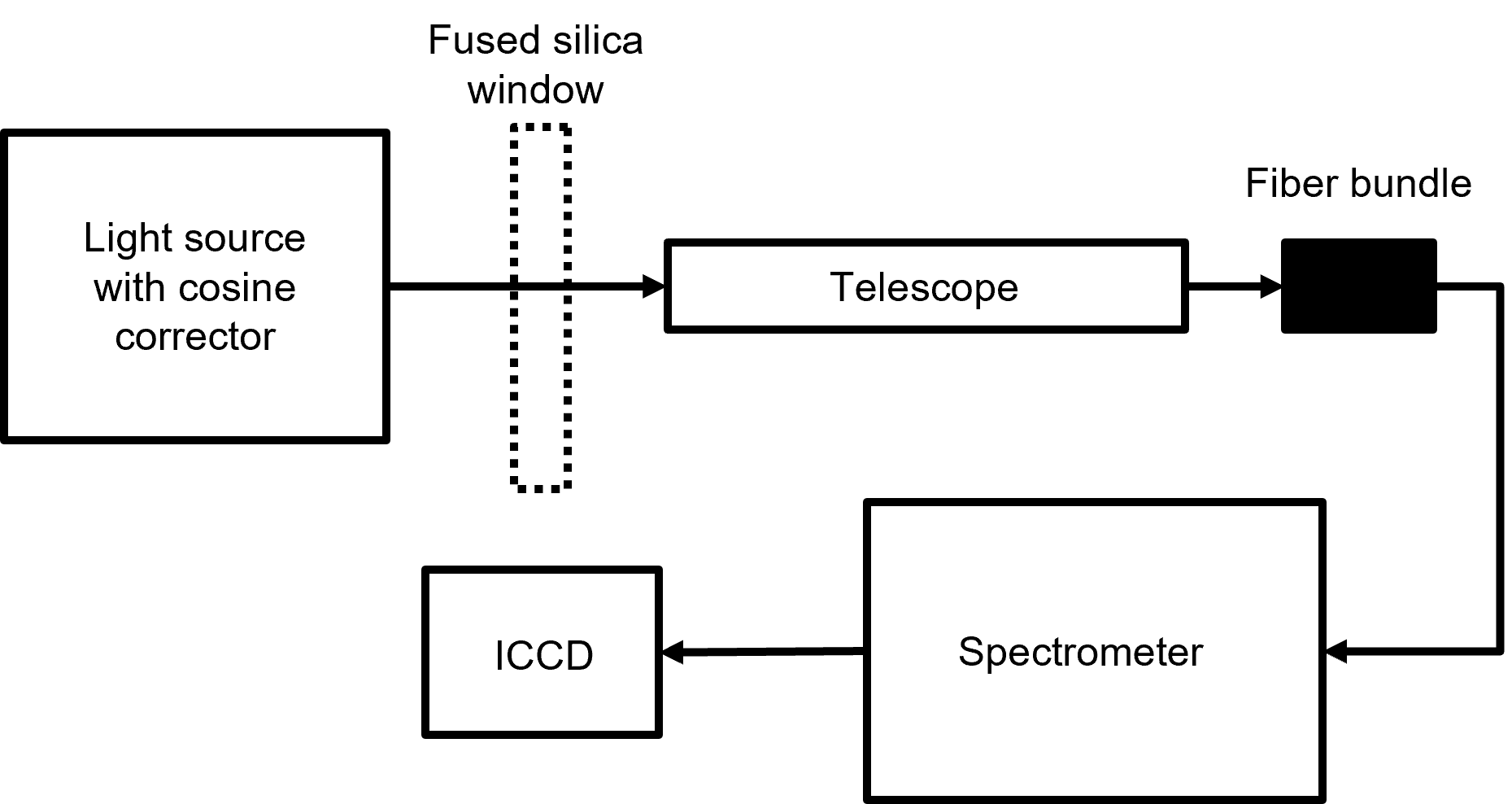}
\caption{\label{fig:calsetup} Absolute calibration setup for S/XB measurements. The output of the light source passes through a fused silica window and is focused by a telecentric telescope onto a 20-chord fiber bundle. The distance from the telescope lens to the light source replicates the distance on the experiment to the Z-pinch plasma. The absolute intensity response of the system is recorded on the ICCD for the known output of the light source.}
\end{figure}

\begin{figure}[h]
\includegraphics[scale=0.6]{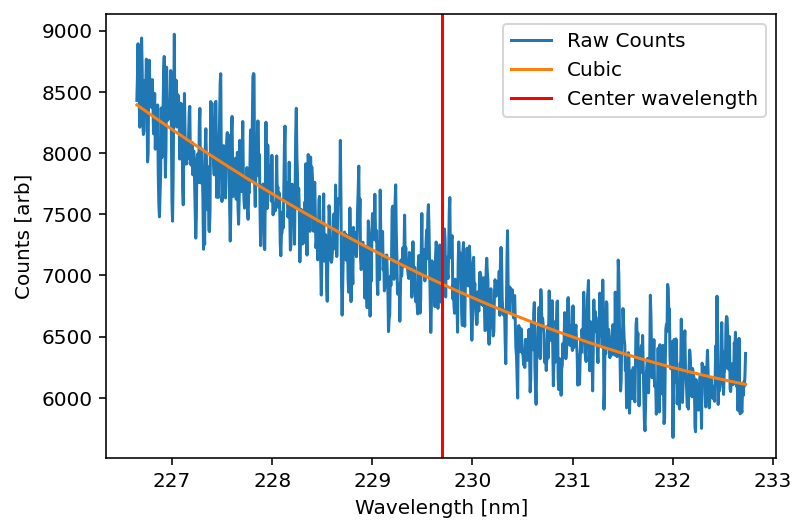}
\caption{\label{fig:cal10} Light source intensity centered at 229.7~nm for chord 10. A cubic function was used to fit the intensity data. The value of the cubic function at the center wavelength is the intensity corresponding to the known absolute irradiance output of the light source.}
\end{figure}

The S/XB method requires measurement of the photon flux of C-III (photons m$^{-2}$s$^{-1}$sr$^{-1}$). To achieve this, the absolute spectral response of the system was calibrated to a light source that outputs a known quantity of light at the desired wavelength. The intensity measured during the experiment was then converted to a photon flux. This calibration was performed individually for all 20 chords in order to account for transmission losses along each optical path. Calibration was done using an Ocean Optics DH-3 Plus light source, which delivers a spectral irradiance of 6.35 $\mu$W/cm$^2$/nm at 229.7 nm. The light source output was cosine-corrected with the CC-3-UV-T diffusive element to ensure isotropic emission. Figure~\ref{fig:calsetup} shows a schematic of the calibration setup. The distance between the light source and the telescope lens replicates the distance between the electrode and the telescope when installed on ZaP-HD. The light source was translated across the optical axis until the intensity was maximized at each of the 20 chords. An acquisition was obtained on the ICCD at each of these positions, which were combined into a single spectrum. The maximum ICCD exposure time of 21 seconds was necessary to gather sufficient light intensity. The signals from each chord were binned and averaged in a similar manner to previous spectroscopy applications. \cite{DenHartog_2001_RSI, Golingo_RSI_2003, Forbes_2020_RSI} For each chord, the intensity variation with wavelength was fitted with a cubic function. The function’s value at 229.7 nm gives the counts corresponding to the absolute irradiance of the light source at that wavelength. This is illustrated in Fig.~\ref{fig:cal10} for chord 10. The resulting calibration data is an array of 20 values for the counts that correspond to the known irradiance output of the light source.
\subsection{\label{sec:calcflux}Calculation of the photon flux}
During operation of the ZaP-HD experiment, the intensity of the C-III emission was recorded over a gate duration of 15~ns. Using the intensity values obtained in the calibration, the absolute irradiance measured during the experiment can be calculated with the expression
\begin{equation}
    E_{exp} = (\frac{C_{exp}}{\tau_{exp}})(\frac{\tau_{cal}}{C_{cal}})E_{cal},
    \label{abscal}
\end{equation}
where the symbol $E$ denotes the absolute irradiance, $C$ is the recorded emission intensity in counts, and $\tau$ is the gate duration of the ICCD acquisition. The subscript $exp$ denotes quantities measured during operation of the ZaP-HD experiment, and the subscript $cal$ is for values obtained during calibration. To calculate the photon flux, $E_{exp}$ is divided by the photon energy,
\begin{equation}
    I = \frac{E_{exp}}{hc}\lambda,
    \label{photonflux}
\end{equation}
where $h$ is the Planck constant, $c$ is the speed of light, and $\lambda$ is 229.7~nm.

\subsection{\label{sec:SXBsel}Applying plasma parameter profiles for S/XB coefficient selection}

Appropriate selection of the S/XB coefficient is made difficult by the rapidly changing plasma parameters near the electrode, and by the limitations in diagnostic measurements of those plasma parameters. On ZaP-HD, the electron density and temperature of the plasma are obtained by Digital Holographic Interferometry (DHI), \cite{Ross2016RSI} which provides radial profiles of the density and temperature over a span of axial locations at a single point in time. The profiles are interpolated to extract a particular density and temperature value at the position of the chords as shown in Fig.~\ref{fig:chord_position}. Interpolation of the dataset in Fig.~\ref{sxbplot} at these plasma parameters gives the corresponding S/XB coefficients. Profiles of the number density obtained from DHI measurements are shown in Fig.~\ref{fig:dhi_density}(a) for the axial location $z$~=~8~cm in the Assembly Region. The number density peaks on axis at 2 $\times$ 10$^{23}$ m$^{-3}$.

The electron temperature plotted in Fig.~\ref{fig:dhi_density}(b) is obtained by the equilibrium analysis described in Ref.~\onlinecite{Shumlak2017}. An abbreviated procedure is provided here. The Z-pinch plasma current is calculated by integrating Amp\`ere's Law and incorporating magnetic field probe data at the axial location of the density measurement. The magnetic field profile is then calculated by integrating the electron number density profile. The magnetic field profile is used in the integration of the radial force balance equation, Eq.~(\ref{radeq}), to calculate the electron temperature profile. The integration is carried out to the pinch radius $a$, beyond which the electron temperature drops to zero. The electron temperature peaks on axis at 670~eV. The characteristic radius $a = 3$~mm from Ref.~\onlinecite{Shumlak2017} is maintained. It is important to note that the DHI measurements provide profiles of the Z-pinch plasma column, while the S/XB measurements are taken where the plasma contacts the electrode. This contact forces the plasma outward along the electrode, which expands the plasma parameter profiles. Therefore, direct application of the density profile from DHI should be interpreted as the upper limit of the plasma parameters used to determine the S/XB coefficients. Modification of this density profile to account for this expansion will be discussed in Sec.~\ref{sec:prescribedprof}.

\subsection{\label{sec:error}Error quantification}

\begin{figure}[h]
\includegraphics[scale=0.57]{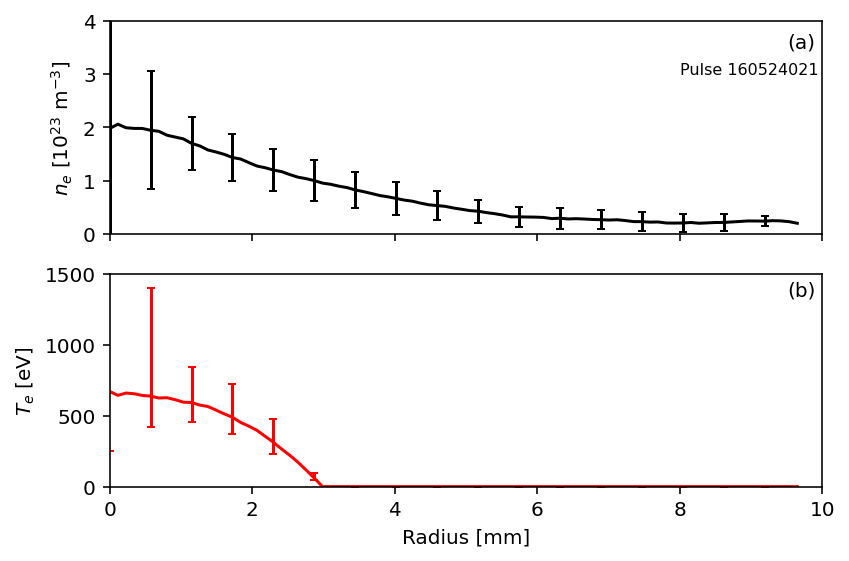}
\caption{\label{fig:dhi_density}Radial profiles of the (a) electron number density and (b) electron temperature obtained from interferometry measurements \cite{Ross2016RSI} of the ZaP-HD Z-pinch plasma. This analysis uses the characteristic pinch radius of 3 mm, beyond which the electron temperature goes to zero. Error bars are shown at sample locations.}
\end{figure}

The error bars for the number density plotted in Fig.~\ref{fig:dhi_density}(a) correspond to the difference between assumed and reconstructed line-integrated density profiles that have been propagated through the same Abel inversion process that provides the number density values. This process is described in further detail in Ref.~\onlinecite{Ross2016RSI}. To propagate the uncertainty in number density, $\delta_{n}$, to the electron temperature, the upper and lower limit of the density defined by the error bars is propagated through the analysis described in Sec.~\ref{sec:SXBsel}. The error bars in Fig.~\ref{fig:dhi_density}(b) represent the difference between these propagated temperature profiles. The range in the S/XB coefficients is then found by interpolation of the data set in Fig.~\ref{sxbplot} for the values of the density and temperature uncertainties to give $\delta_{SXB}$. The other source of uncertainty in the flux measurement is from the emission intensity, which is assumed to follow Poisson statistics. The uncertainty in the counts is therefore $\delta_{C} = \sqrt{C_{exp}}$, which is propagated through Eqs.~(\ref{abscal}) and (\ref{photonflux}) to obtain the uncertainty in the photon flux, $\delta_{I}$. The overall uncertainty in the eroded flux measurement, $\delta_{\Gamma}$, is calculated with 
\begin{equation}
    \frac{\delta_{\Gamma}}{\Gamma} = \sqrt{\left( \frac{\delta_{SXB}}{S/XB} \right)^2 + \left( \frac{\delta_{I}}{I} \right)^2}.
    \label{fluxerr}
\end{equation}
\section{\label{sec:results}Measurements of eroded carbon flux from C-III intensity}

\begin{figure}[h]
\includegraphics[scale=0.6]{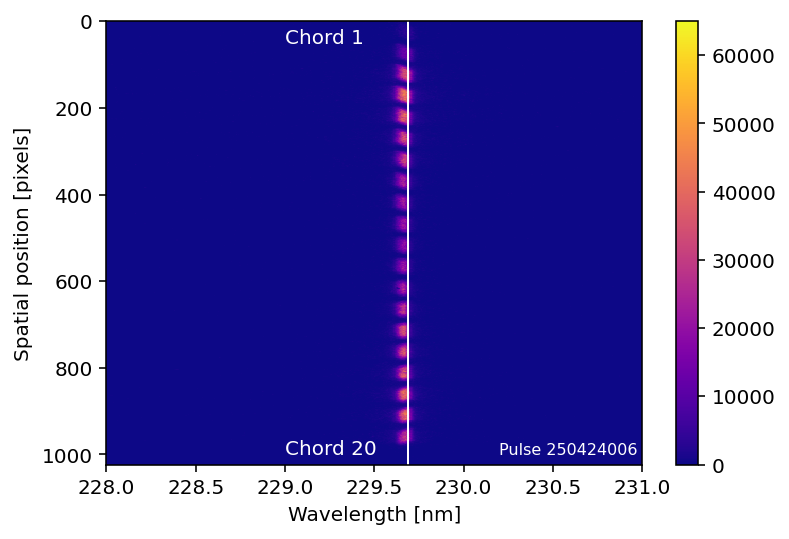}
\caption{\label{fig:spectrum} Chord-integrated line emission of C-III at 229.7 nm. The central chords, which terminate on the electrode, record a lower intensity compared to the outer chords. The solid line is positioned at 229.7~nm for reference. The 20 chords are arranged as shown in Fig.~\ref{fig:chord_position}, with the telescope orientation as shown in Fig.~\ref{fig:telescope_position}.}
\end{figure}

The S/XB diagnostic recorded C-III emission on the ZaP-HD device for pure hydrogen Z-pinch plasmas. A raw spectrum showing the emission captured at 229.7~nm is shown in Fig.~\ref{fig:spectrum}. The solid line is positioned at 229.7~nm for reference. The spatial position corresponds to the $x$ dimension in Fig.~\ref{fig:chord_position}. The ICCD was triggered at 50~$\mu$s to capture C-III emission coincident with the peak pinch current, and therefore the peak density and temperature. This timing also aligns with the DHI measurement. The intensity of emission required a gate width of 15~ns to avoid saturation of the sensor. The emission intensity decreases close to the axis where the chords terminate on the electrode, while more C-III emission is captured along the outer chords that have a longer line of sight. This supports the use of the outer chords as a background subtraction described in Sec.~\ref{sec:sxbsetup} to isolate the C-III emission local to the surface.

\begin{figure}[h]
\includegraphics[scale=0.6]{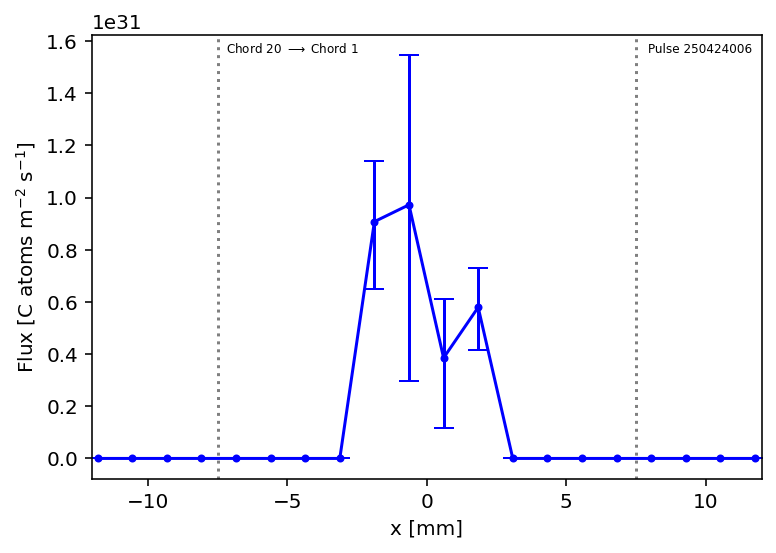}
\caption{\label{fig:cflux}The eroded flux of carbon atoms as measured by the S/XB spectroscopy diagnostic. The S/XB coefficients at each impact parameter $x$ are determined using the radial density and temperature profiles shown in Fig.~\ref{fig:dhi_density}. Beyond the assumed pinch radius of 3 mm, the electron temperature profile drops to zero, therefore S/XB coefficients are only assigned to the central four chords for calculation of the eroded flux.}
\end{figure}

Applying the methods described in Sec.~\ref{sec:dataprocessing} provides the inferred carbon erosion fluxes shown in Fig.~\ref{fig:cflux}.  Values of the flux are only calculated for chord positions within the pinch radius of 3~mm, which have a non-zero electron temperature according to the profile specified in Fig.~\ref{fig:dhi_density}(b). The vertical dotted lines indicate the edge of the electrode surface. The peak eroded flux is $9.7~\times~10^{30}$~atoms~ m$^{-2}$s$^{-1}$ at -0.6~mm offset from the experiment axis (chord 11). However, the error bars indicate that the peak may occur further off-axis at 2~mm. Large errors in the eroded flux closer to the axis is likely due to corresponding large errors in density and temperature as shown in Fig.~\ref{fig:dhi_density}. 

\subsection{\label{sec:compareflux}Comparison to theoretical sputtering and sublimation fluxes}

This section compares the measured carbon erosion fluxes to the theoretical values from physical sputtering and sublimation, which provides some insight into the physical mechanisms involved in electrode erosion. Physical sputtering arises from the impact of energetic plasma ions on the electrode surface. A neutral carbon atom is sputtered if the energy transferred in the collision exceeds the binding energy between carbon atoms in the solid lattice. Sublimation arises from the absorption of heat which provides sufficient energy for carbon atoms in the lattice to escape as a vapor. Modeling  \cite{Thompson_2023_POP} of the electrode response on the FuZE SFS Z-pinch device \cite{Zhang_2019_PRL} incorporates both of these processes in its erosion model, and predicts the surface temperature to far exceed the sublimation temperature of graphite.

\begin{figure}[h]
\includegraphics[scale=0.6]{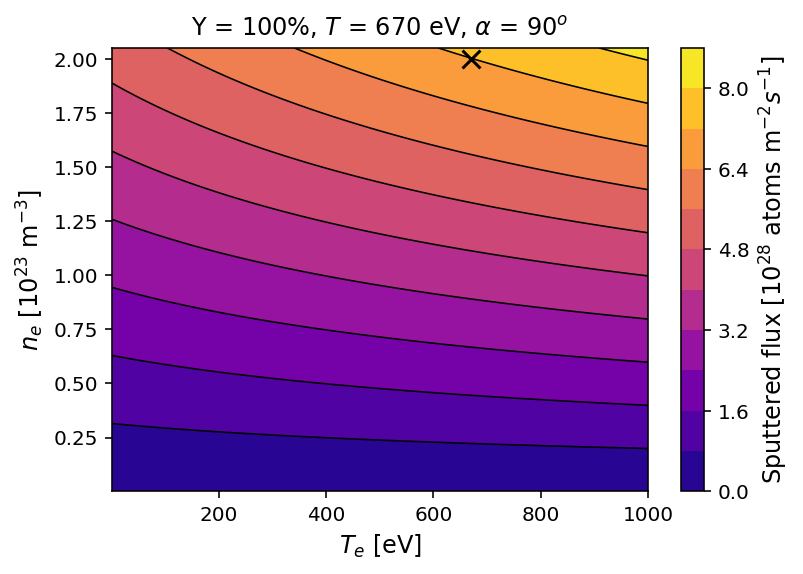}
\caption{\label{fig:physsput} Theoretical values for the eroded carbon flux due to physical sputtering by hydrogen ions, assuming a sputtering yield of 100$\%$, $T = T_i = 670$~eV, and normal incidence of impacting ions. The value indicated by the marker is used for comparison to measured erosion fluxes.}
\end{figure}

Since ZaP-HD operates with a hydrogen plasma, $H^+$ is assumed to be the dominant ion species. The following calculations assume that incident H$^+$ ions impact at normal incidence with a sputtering yield of $100\%$. This provides the upper limit of sputtering that can occur for H$^+$ ions on carbon. From the analysis in Ref.~\onlinecite{Shumlak2017}, the assumption of $T_i = T_e = T$ is applied, where $T = 670$ eV from Fig.~\ref{fig:dhi_density}(b). The sputtered carbon flux is calculated with the following expression, adapted from Ref.~\onlinecite{Naujoks_1996}:
\begin{equation}
    \Gamma_{sp} = n_e c_s sin(\alpha) Y^*.
    \label{physsputeq}
\end{equation}
Here, $n_e$ is the electron number density, and $c_s$ is the sound speed expressed as $[k(T_e + T_i)/m_i]^{1/2}$. For normal incidence, $\alpha = 90^o$ and $Y^*$ is the effective yield defined as
\begin{equation}
    Y^* = Y_{H^+\rightarrow C}f_{H^+},
\end{equation}
where $Y_{H^+\rightarrow C} = 1$ represents 100$\%$ yield for physical sputtering of carbon by hydrogen ions, and $f_{H^+}$ is the fraction of the plasma composed of hydrogen, in this case equal to 1. The sputtered flux is calculated for the same range of $n_e$ and $T_e$ as Fig.~\ref{fig:sxbcoeffs}. The resulting values for $\Gamma_{sp}$ are plotted in Fig.~\ref{fig:physsput}, with the expected peak value for ZaP-HD indicated by the marker. This sputtered flux is $7.2\times10^{28}$ atoms m$^{-2}$s$^{-1}$, almost three orders of magnitude smaller than the peak erosion flux shown in Fig.~\ref{fig:cflux}. 

To calculate the sublimation flux of carbon, the entirety of the heat flux from the plasma reaching the electrode surface is assumed to contribute to heating and sublimation. Magnetic field effects, which may reduce the flux to the surface, are neglected. The particle flux at the sheath edge is assumed to be the particle flux at the solid surface. Similarly to the calculation of the sputtered flux, these assumptions provide the upper limit to the calculated flux. Following the analysis in Ref.~\onlinecite{stangeby_2000}, the heat flux to the surface is approximated by
\begin{equation}
    q = (2kT_i + \mid eV_{sheath}\mid)\Gamma_{se},
    \label{heatfluxeq}
\end{equation}
where $T_i$ is the ion temperature, $V_{sheath}$ is the voltage drop through the sheath, and $\Gamma_{se}$ is the particle flux at the sheath edge. For this analysis, $V_{sheath}$ is assumed to be the applied voltage to the electrode, approximated from voltage measurements on ZaP-HD as 4.8~kV. The particle flux is expressed as
\begin{equation}
    \Gamma_{se} = n_{se}c_{s} \approx \frac{1}{2}n_0[k(T_e + T_i)/m_i]^{1/2}, 
    \label{partfluxeq}
\end{equation}
where $n_{se}$ is the electron number density at the sheath edge. The calculation uses the peak density in Fig.~\ref{fig:dhi_density}(a) as $n_0$, the same temperature values used in Eq.~\ref{physsputeq}, and the ion mass of hydrogen as $m_i$. To obtain the flux of sublimated carbon, the heat flux calculated in Eq.~(\ref{heatfluxeq}) is divided by the surface binding energy of 7.4~eV, which is the heat of sublimation of graphite. \cite{dasent1982inorganic} The result is a theoretical sublimated flux of $3 \times10^{31}$~atoms~m$^{-2}$s$^{-1}$, which is comparable to the values in Fig.~\ref{fig:cflux}. This suggests that ionization of sublimated carbon atoms is responsible for the bulk of the C-III emission recorded by the S/XB diagnostic, while the sputtered carbon flux has a very small contribution to the measurement. This is consistent with the evaporative processes that dominate electrode erosion in arc plasmas. \cite{Nemchinsky_2014} A much larger discrepancy between sputtered and sublimation fluxes of five orders of magnitude was observed in the FuZE simulations. \cite{Thompson_2023_POP} This larger discrepancy is likely due to incorporation of redeposition effects which reduce the sputtered flux. Regardless, such a discrepancy between these two physical processes encourages further analysis of these S/XB measurements. For example, consideration of the relative energies of sublimated carbon compared to sputtered carbon may provide insight on the likelihood of local redeposition and ultimately net erosion of the electrode.

\subsection{\label{sec:prescribedprof}Expanded density profiles for S/XB calculations}

\begin{figure}[h]
\includegraphics[scale=0.63]{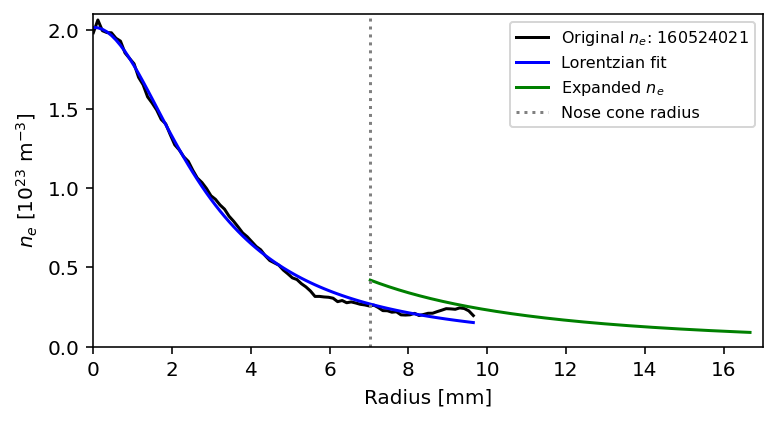}
\caption{\label{fig:expanded}Original (black) and expanded (green) electron number density profiles using data from DHI. A Lorentzian fit (blue) is used to approximate the original profile and to generate the expanded profile. The linear density of both profiles is conserved according to Eq.~(\ref{lineardenscons}). The dotted gray line represents the electrode radius at the nose cone. The peak density is reduced by a factor of four in the expanded profile.}
\end{figure}

\begin{figure}[h]
\includegraphics[scale=0.6]{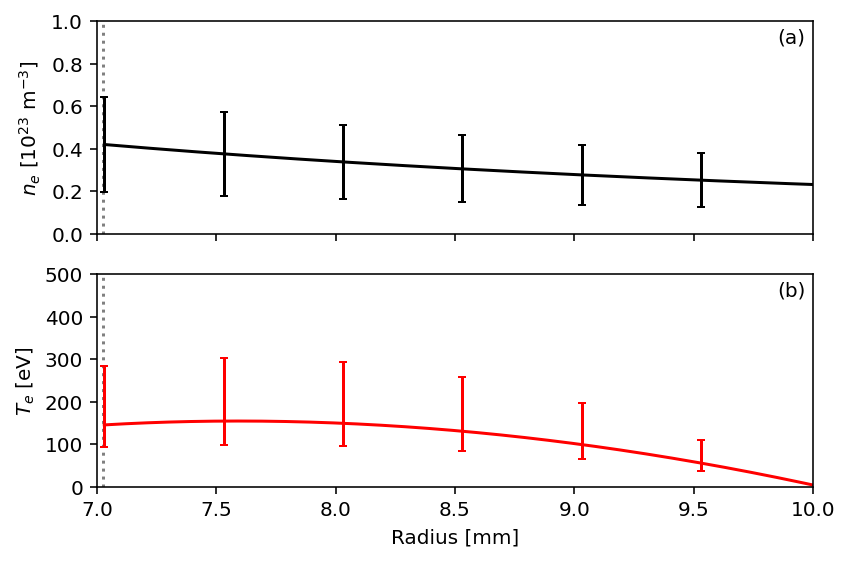}
\caption{\label{fig:expandedprofiles} Expanded radial profiles of the (a) electron number density and (b) electron temperature, based on original profiles in Fig.~\ref{fig:dhi_density}. Expanded profiles begin at the radius of the electrode at the nose cone, shown as the dotted gray line, which accounts for plasma contact with the electrode. The peak electron temperature is reduced by about a factor of five compared to the original profile, to 145~eV.}
\end{figure}

The results shown in Fig.~\ref{fig:cflux} apply the measured profiles of the core plasma to the S/XB measurement locations on the electrode. In reality, the geometry of the electrode forces the plasma outwards at the location of the S/XB measurement. For a constant linear density, the profiles should expand, reducing the density at the electrode. At present, local measurements of the plasma parameters on the electrode are not available on ZaP-HD. However, the measured profiles can be modified to account for this expansion. The Z-pinch mass, and therefore the linear density, is held constant, and the expanded profile redistributes the conserved mass. To implement this, the number density measured by DHI in Fig.~\ref{fig:dhi_density}(a) is integrated using Eq.~(\ref{lineardensityeq}) to give the linear density $N$. Using the electrode radius as the lower limit of integration and imposing an equality for the linear density,
\begin{equation}
    N = \int_0^\infty 2\pi n_{1}(r)rdr =  \int_{r_{nc}}^\infty 2\pi n_{2}(r) rdr,
    \label{lineardenscons}
\end{equation}
where $n_1$ is the original number density profile, $n_2$ is the expanded number density profile, and $r_{nc}$ is the radius of the electrode on the nose cone where the S/XB measurement is taken. The calculated linear density is $N=1.26$~$\times$~10$^{19}$~m$^{-1}$. A comparison of the profiles is illustrated in Fig.~\ref{fig:expanded}. The original density profile $n_1$ is shown in black. A Lorentzian fit provides a good approximation of $n_1$, shown in blue. Therefore, a Lorentzian is used to generate the expanded profile, shown in green, with amplitude and width parameters such that the linear density $N$ is conserved. The expanded profile starts at $r_{nc} = 7$~mm, shown as a dotted gray line. Applying the same equilibrium analysis from Ref.~\onlinecite{Shumlak2017} gives the electron temperature, illustrated along with the expanded density profile in Fig.~\ref{fig:expandedprofiles}. The radius of $a = 3$~mm has been maintained. The peak density of the expanded profile decreases to 4.2~$\times$~10$^{22}$~m$^{-3}$, while the peak electron temperature decreases to 145~eV. The error bars are calculated by performing the same expansion procedure on density profiles defined by the error bars of the original profiles. For the same linear density, a reduced number density and temperature correspond to a reduced pinch current as shown by the scaling relations \cite{Shumlak2012_FST, Shumlak2017, Shumlak2020} from Sec.~\ref{sec:zaphdsec}.

\begin{figure}[h]
\includegraphics[scale=0.6]{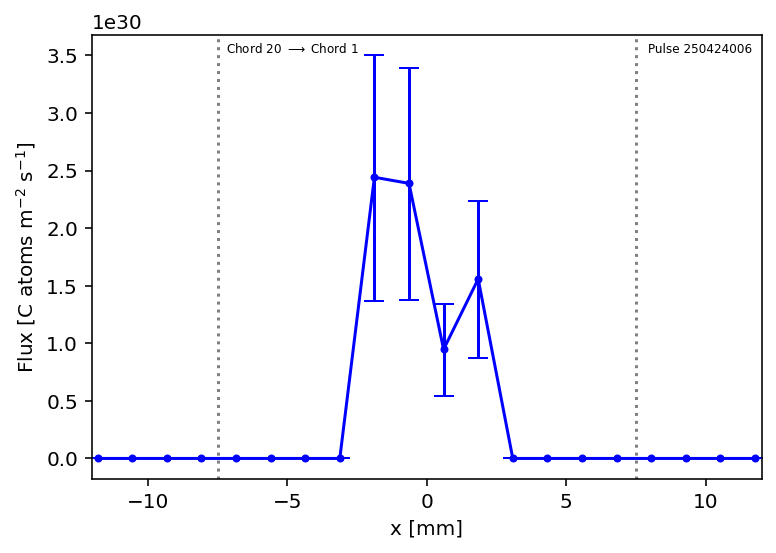}
\caption{\label{fig:cflux2} The eroded flux of carbon atoms as measured by the S/XB spectroscopy diagnostic. The S/XB coefficients at each impact parameter $x$ are determined using the expanded radial density and temperature profiles from Fig.~\ref{fig:expandedprofiles}. Fluxes are reduced by about a factor of four compared to those calculated using the profiles in Fig.~\ref{fig:dhi_density}.}
\end{figure}

To calculate the erosion flux with the expanded profile, the S/XB coefficients are found using the density and temperature values at $r_{nc}$ since each chord terminates on the nose cone at the same radius. The results are plotted in Fig.~\ref{fig:cflux2}. The peak erosion flux is approximately a factor of four smaller than that calculated with the original profiles, decreasing to $2.4~\times~10^{30}$~atoms~m$^{-2}$s$^{-1}$. Despite this decrease, it remains too large for physical sputtering to account entirely for the measurement, especially considering the assumptions made to maximize the theoretical sputtered flux. The erosion fluxes in Fig.~\ref{fig:cflux2} are well below the maximum theoretical sublimation flux. These results continue to support the theory that sublimation is responsible for the majority of the C-III emission recorded.

\section{\label{sec:conc}Conclusions}
The S/XB, or ionizations per photon, method has been implemented on the ZaP-HD experiment to measure the flux of eroded carbon from the graphite electrode via UV emission of C-III ions at 229.7 nm. Absolute intensity calibration of the spectroscopy system enables the use of S/XB coefficients to infer the eroded flux of carbon atoms from the photon flux of C-III emission. Selection of S/XB coefficients depends on the local plasma density and temperature, which influence the ionization and excitation rates of the C-III species. Radial profiles of the electron number density and electron temperature from DHI measurements of the Z-pinch plasma are applied to spectroscopy chord locations on the electrode, which limits the measurement to four chord locations that are within the specified pinch radius of 3 mm. Uncertainty in the S/XB coefficients comes from the error in the DHI measurement of number density, which is used to calculate the variation in the electron temperature through equilibrium analysis. This is then combined with the statistical uncertainty in the measured intensity to obtain the uncertainty in the erosion flux measurement.

The measured carbon erosion flux peaks at $9.7~\times~10^{30}$~atoms~ m$^{-2}$s$^{-1}$, nearly three orders of magnitude larger than the theoretical sputtered flux, but within an order of magnitude of the theoretical sublimation flux. By conserving the linear density, the number density profile is modified to account for expansion of the plasma as it contacts the electrode, which decreases the local plasma parameters used to determine the S/XB coefficients. The calculated erosion fluxes from this expanded profile are about four times smaller than those calculated with the original density profile, but remain much larger than the predicted sputtered flux. These results suggest that sublimation is primarily responsible for the erosion process, at least in terms of initial removal of carbon from the electrode. This dominance of sublimation over sputtering implies a large population of neutrals entering the plasma with energy characteristic of the electrode surface temperature. A comparison to the energy of sputtered neutrals could have implications on redeposition and net erosion, which warrants further analysis. 

The S/XB diagnostic forms one component of a broader physics study on ZaP-HD that includes measurements of net mass loss, surface temperature, and microscopic surface morphology changes over various plasma exposure conditions that enables a more complete understanding of the erosion behavior of the electrode material. This understanding can guide configurational and operational design choices in future fusion devices, especially with regard to durability of electrodes in direct contact with the plasma.

\begin{acknowledgments}
The information, data, or work presented herein was funded in part by the National Nuclear Security Administration under Grant No. DE-NA0001860.
\end{acknowledgments}

\section*{Author Declarations}
\subsection*{Conflict of Interest}

The authors have no conflicts to disclose.

\subsection*{Author Contributions}
\noindent \textbf{Amierul Aqil Khairi:} Conceptualization (equal); Formal analysis (lead); Investigation (lead); Methodology (equal); Writing - original draft (lead); Writing - review \& editing (equal). \textbf{Elyse Lian:} Formal analysis (supporting); Investigation (supporting); Methodology (supporting); Writing - review \& editing (supporting). \textbf{Uri Shumlak:} Conceptualization (equal); Funding acquisition (lead); Methodology (equal); Supervision (lead); Writing - review \& editing (equal).

\section*{Data Availability Statement}

The data that support the findings of this study are available from the corresponding author upon reasonable request.



\nocite{*}
\bibliography{aipsamp}

\providecommand{\noopsort}[1]{}\providecommand{\singleletter}[1]{#1}%
\begin{thebibliography}{41}%
\makeatletter
\providecommand \@ifxundefined [1]{%
 \@ifx{#1\undefined}
}%
\providecommand \@ifnum [1]{%
 \ifnum #1\expandafter \@firstoftwo
 \else \expandafter \@secondoftwo
 \fi
}%
\providecommand \@ifx [1]{%
 \ifx #1\expandafter \@firstoftwo
 \else \expandafter \@secondoftwo
 \fi
}%
\providecommand \natexlab [1]{#1}%
\providecommand \enquote  [1]{``#1''}%
\providecommand \bibnamefont  [1]{#1}%
\providecommand \bibfnamefont [1]{#1}%
\providecommand \citenamefont [1]{#1}%
\providecommand \href@noop [0]{\@secondoftwo}%
\providecommand \href [0]{\begingroup \@sanitize@url \@href}%
\providecommand \@href[1]{\@@startlink{#1}\@@href}%
\providecommand \@@href[1]{\endgroup#1\@@endlink}%
\providecommand \@sanitize@url [0]{\catcode `\\12\catcode `\$12\catcode `\&12\catcode `\#12\catcode `\^12\catcode `\_12\catcode `\%12\relax}%
\providecommand \@@startlink[1]{}%
\providecommand \@@endlink[0]{}%
\providecommand \url  [0]{\begingroup\@sanitize@url \@url }%
\providecommand \@url [1]{\endgroup\@href {#1}{\urlprefix }}%
\providecommand \urlprefix  [0]{URL }%
\providecommand \Eprint [0]{\href }%
\providecommand \doibase [0]{http://dx.doi.org/}%
\providecommand \selectlanguage [0]{\@gobble}%
\providecommand \bibinfo  [0]{\@secondoftwo}%
\providecommand \bibfield  [0]{\@secondoftwo}%
\providecommand \translation [1]{[#1]}%
\providecommand \BibitemOpen [0]{}%
\providecommand \bibitemStop [0]{}%
\providecommand \bibitemNoStop [0]{.\EOS\space}%
\providecommand \EOS [0]{\spacefactor3000\relax}%
\providecommand \BibitemShut  [1]{\csname bibitem#1\endcsname}%
\let\auto@bib@innerbib\@empty
\bibitem [{\citenamefont {Zinkle}(2013)}]{Zinkle2013}%
  \BibitemOpen
  \bibfield  {author} {\bibinfo {author} {\bibfnamefont {S.~J.}\ \bibnamefont {Zinkle}},\ }\bibfield  {title} {\enquote {\bibinfo {title} {Challenges in developing materials for fusion technology - past, present and future},}\ }\href {\doibase 10.13182/FST13-631} {\bibfield  {journal} {\bibinfo  {journal} {Fusion Science and Technology}\ }\textbf {\bibinfo {volume} {64}},\ \bibinfo {pages} {65--75} (\bibinfo {year} {2013})},\ \Eprint {http://arxiv.org/abs/https://doi.org/10.13182/FST13-631} {https://doi.org/10.13182/FST13-631} \BibitemShut {NoStop}%
\bibitem [{\citenamefont {Linsmeier}\ \emph {et~al.}(2017{\natexlab{a}})\citenamefont {Linsmeier}, \citenamefont {Rieth}, \citenamefont {Aktaa}, \citenamefont {Chikada}, \citenamefont {Hoffmann}, \citenamefont {Hoffmann}, \citenamefont {Houben}, \citenamefont {Kurishita}, \citenamefont {Jin}, \citenamefont {Li}, \citenamefont {Litnovsky}, \citenamefont {Matsuo}, \citenamefont {von Müller}, \citenamefont {Nikolic}, \citenamefont {Palacios}, \citenamefont {Pippan}, \citenamefont {Qu}, \citenamefont {Reiser}, \citenamefont {Riesch}, \citenamefont {Shikama}, \citenamefont {Stieglitz}, \citenamefont {Weber}, \citenamefont {Wurster}, \citenamefont {You},\ and\ \citenamefont {Zhou}}]{Linsmeier_2017}%
  \BibitemOpen
  \bibfield  {author} {\bibinfo {author} {\bibfnamefont {C.}~\bibnamefont {Linsmeier}}, \bibinfo {author} {\bibfnamefont {M.}~\bibnamefont {Rieth}}, \bibinfo {author} {\bibfnamefont {J.}~\bibnamefont {Aktaa}}, \bibinfo {author} {\bibfnamefont {T.}~\bibnamefont {Chikada}}, \bibinfo {author} {\bibfnamefont {A.}~\bibnamefont {Hoffmann}}, \bibinfo {author} {\bibfnamefont {J.}~\bibnamefont {Hoffmann}}, \bibinfo {author} {\bibfnamefont {A.}~\bibnamefont {Houben}}, \bibinfo {author} {\bibfnamefont {H.}~\bibnamefont {Kurishita}}, \bibinfo {author} {\bibfnamefont {X.}~\bibnamefont {Jin}}, \bibinfo {author} {\bibfnamefont {M.}~\bibnamefont {Li}}, \bibinfo {author} {\bibfnamefont {A.}~\bibnamefont {Litnovsky}}, \bibinfo {author} {\bibfnamefont {S.}~\bibnamefont {Matsuo}}, \bibinfo {author} {\bibfnamefont {A.}~\bibnamefont {von Müller}}, \bibinfo {author} {\bibfnamefont {V.}~\bibnamefont {Nikolic}}, \bibinfo {author} {\bibfnamefont {T.}~\bibnamefont {Palacios}}, \bibinfo {author} {\bibfnamefont {R.}~\bibnamefont
  {Pippan}}, \bibinfo {author} {\bibfnamefont {D.}~\bibnamefont {Qu}}, \bibinfo {author} {\bibfnamefont {J.}~\bibnamefont {Reiser}}, \bibinfo {author} {\bibfnamefont {J.}~\bibnamefont {Riesch}}, \bibinfo {author} {\bibfnamefont {T.}~\bibnamefont {Shikama}}, \bibinfo {author} {\bibfnamefont {R.}~\bibnamefont {Stieglitz}}, \bibinfo {author} {\bibfnamefont {T.}~\bibnamefont {Weber}}, \bibinfo {author} {\bibfnamefont {S.}~\bibnamefont {Wurster}}, \bibinfo {author} {\bibfnamefont {J.-H.}\ \bibnamefont {You}}, \ and\ \bibinfo {author} {\bibfnamefont {Z.}~\bibnamefont {Zhou}},\ }\bibfield  {title} {\enquote {\bibinfo {title} {Development of advanced high heat flux and plasma-facing materials},}\ }\href {\doibase 10.1088/1741-4326/aa6f71} {\bibfield  {journal} {\bibinfo  {journal} {Nuclear Fusion}\ }\textbf {\bibinfo {volume} {57}},\ \bibinfo {pages} {092007} (\bibinfo {year} {2017}{\natexlab{a}})},\ \Eprint {http://arxiv.org/abs/https://doi.org/10.1088/1741-4326/aa6f71} {https://doi.org/10.1088/1741-4326/aa6f71}
  \BibitemShut {NoStop}%
\bibitem [{\citenamefont {Linke}\ \emph {et~al.}(2019)\citenamefont {Linke}, \citenamefont {Du}, \citenamefont {Loewenhoff}, \citenamefont {Pintsuk}, \citenamefont {Spilker}, \citenamefont {Steudel},\ and\ \citenamefont {Wirtz}}]{Linke_2019_MRE}%
  \BibitemOpen
  \bibfield  {author} {\bibinfo {author} {\bibfnamefont {J.}~\bibnamefont {Linke}}, \bibinfo {author} {\bibfnamefont {J.}~\bibnamefont {Du}}, \bibinfo {author} {\bibfnamefont {T.}~\bibnamefont {Loewenhoff}}, \bibinfo {author} {\bibfnamefont {G.}~\bibnamefont {Pintsuk}}, \bibinfo {author} {\bibfnamefont {B.}~\bibnamefont {Spilker}}, \bibinfo {author} {\bibfnamefont {I.}~\bibnamefont {Steudel}}, \ and\ \bibinfo {author} {\bibfnamefont {M.}~\bibnamefont {Wirtz}},\ }\bibfield  {title} {\enquote {\bibinfo {title} {Challenges for plasma-facing components in nuclear fusion},}\ }\href {\doibase 10.1063/1.5090100} {\bibfield  {journal} {\bibinfo  {journal} {Matter and Radiation at Extremes}\ }\textbf {\bibinfo {volume} {4}},\ \bibinfo {pages} {056201} (\bibinfo {year} {2019})},\ \Eprint {http://arxiv.org/abs/https://doi.org/10.1063/1.5090100} {https://doi.org/10.1063/1.5090100} \BibitemShut {NoStop}%
\bibitem [{\citenamefont {Shumlak}(2020)}]{Shumlak2020}%
  \BibitemOpen
  \bibfield  {author} {\bibinfo {author} {\bibfnamefont {U.}~\bibnamefont {Shumlak}},\ }\bibfield  {title} {\enquote {\bibinfo {title} {Z-pinch fusion},}\ }\href {\doibase 10.1063/5.0004228} {\bibfield  {journal} {\bibinfo  {journal} {Journal of Applied Physics}\ }\textbf {\bibinfo {volume} {127}},\ \bibinfo {pages} {200901} (\bibinfo {year} {2020})},\ \Eprint {http://arxiv.org/abs/https://doi.org/10.1063/5.0004228} {https://doi.org/10.1063/5.0004228} \BibitemShut {NoStop}%
\bibitem [{\citenamefont {Ikeda}(2007)}]{Ikeda_2007}%
  \BibitemOpen
  \bibfield  {author} {\bibinfo {author} {\bibfnamefont {K.}~\bibnamefont {Ikeda}},\ }\bibfield  {title} {\enquote {\bibinfo {title} {Progress in the iter physics basis},}\ }\href {\doibase 10.1088/0029-5515/47/6/E01} {\bibfield  {journal} {\bibinfo  {journal} {Nuclear Fusion}\ }\textbf {\bibinfo {volume} {47}},\ \bibinfo {pages} {E01} (\bibinfo {year} {2007})},\ \Eprint {http://arxiv.org/abs/https://dx.doi.org/10.1088/0029-5515/47/6/E01} {https://dx.doi.org/10.1088/0029-5515/47/6/E01} \BibitemShut {NoStop}%
\bibitem [{\citenamefont {Kreter}(2011)}]{Kreter2011}%
  \BibitemOpen
  \bibfield  {author} {\bibinfo {author} {\bibfnamefont {A.}~\bibnamefont {Kreter}},\ }\bibfield  {title} {\enquote {\bibinfo {title} {Reactor-relevant plasma-material interaction studies in linear plasma devices},}\ }\href {\doibase 10.13182/FST59-1T-51} {\bibfield  {journal} {\bibinfo  {journal} {Fusion Science and Technology}\ }\textbf {\bibinfo {volume} {59}},\ \bibinfo {pages} {51--56} (\bibinfo {year} {2011})},\ \Eprint {http://arxiv.org/abs/https://doi.org/10.13182/FST59-1T-51} {https://doi.org/10.13182/FST59-1T-51} \BibitemShut {NoStop}%
\bibitem [{\citenamefont {Linsmeier}\ \emph {et~al.}(2017{\natexlab{b}})\citenamefont {Linsmeier}, \citenamefont {Unterberg}, \citenamefont {Coenen}, \citenamefont {Doerner}, \citenamefont {Greuner}, \citenamefont {Kreter}, \citenamefont {Linke},\ and\ \citenamefont {Maier}}]{Linsmeier_2017_Facilities}%
  \BibitemOpen
  \bibfield  {author} {\bibinfo {author} {\bibfnamefont {C.}~\bibnamefont {Linsmeier}}, \bibinfo {author} {\bibfnamefont {B.}~\bibnamefont {Unterberg}}, \bibinfo {author} {\bibfnamefont {J.}~\bibnamefont {Coenen}}, \bibinfo {author} {\bibfnamefont {R.}~\bibnamefont {Doerner}}, \bibinfo {author} {\bibfnamefont {H.}~\bibnamefont {Greuner}}, \bibinfo {author} {\bibfnamefont {A.}~\bibnamefont {Kreter}}, \bibinfo {author} {\bibfnamefont {J.}~\bibnamefont {Linke}}, \ and\ \bibinfo {author} {\bibfnamefont {H.}~\bibnamefont {Maier}},\ }\bibfield  {title} {\enquote {\bibinfo {title} {Material testing facilities and programs for plasma-facing component testing},}\ }\href {\doibase 10.1088/1741-4326/aa4feb} {\bibfield  {journal} {\bibinfo  {journal} {Nuclear Fusion}\ }\textbf {\bibinfo {volume} {57}},\ \bibinfo {pages} {092012} (\bibinfo {year} {2017}{\natexlab{b}})},\ \Eprint {http://arxiv.org/abs/https://dx.doi.org/10.1088/1741-4326/aa4feb} {https://dx.doi.org/10.1088/1741-4326/aa4feb} \BibitemShut {NoStop}%
\bibitem [{\citenamefont {Shumlak}\ \emph {et~al.}(2017)\citenamefont {Shumlak}, \citenamefont {Nelson}, \citenamefont {Claveau}, \citenamefont {Forbes}, \citenamefont {Golingo}, \citenamefont {Hughes}, \citenamefont {Oberto}, \citenamefont {Ross},\ and\ \citenamefont {Weber}}]{Shumlak2017}%
  \BibitemOpen
  \bibfield  {author} {\bibinfo {author} {\bibfnamefont {U.}~\bibnamefont {Shumlak}}, \bibinfo {author} {\bibfnamefont {B.~A.}\ \bibnamefont {Nelson}}, \bibinfo {author} {\bibfnamefont {E.~L.}\ \bibnamefont {Claveau}}, \bibinfo {author} {\bibfnamefont {E.~G.}\ \bibnamefont {Forbes}}, \bibinfo {author} {\bibfnamefont {R.~P.}\ \bibnamefont {Golingo}}, \bibinfo {author} {\bibfnamefont {M.~C.}\ \bibnamefont {Hughes}}, \bibinfo {author} {\bibfnamefont {R.~J.}\ \bibnamefont {Oberto}}, \bibinfo {author} {\bibfnamefont {M.~P.}\ \bibnamefont {Ross}}, \ and\ \bibinfo {author} {\bibfnamefont {T.~R.}\ \bibnamefont {Weber}},\ }\bibfield  {title} {\enquote {\bibinfo {title} {Increasing plasma parameters using sheared flow stabilization of a z-pinch},}\ }\href {\doibase 10.1063/1.4977468} {\bibfield  {journal} {\bibinfo  {journal} {Physics of Plasmas}\ }\textbf {\bibinfo {volume} {24}},\ \bibinfo {pages} {055702} (\bibinfo {year} {2017})},\ \Eprint {http://arxiv.org/abs/https://doi.org/10.1063/1.4977468}
  {https://doi.org/10.1063/1.4977468} \BibitemShut {NoStop}%
\bibitem [{\citenamefont {Thompson}\ \emph {et~al.}(2023{\natexlab{a}})\citenamefont {Thompson}, \citenamefont {Levitt}, \citenamefont {Nelson},\ and\ \citenamefont {Shumlak}}]{Thompson_2023_FST}%
  \BibitemOpen
  \bibfield  {author} {\bibinfo {author} {\bibfnamefont {M.}~\bibnamefont {Thompson}}, \bibinfo {author} {\bibfnamefont {B.}~\bibnamefont {Levitt}}, \bibinfo {author} {\bibfnamefont {B.}~\bibnamefont {Nelson}}, \ and\ \bibinfo {author} {\bibfnamefont {U.}~\bibnamefont {Shumlak}},\ }\bibfield  {title} {\enquote {\bibinfo {title} {Engineering paradigms for sheared-flow-stabilized z-pinch fusion energy},}\ }\href {\doibase 10.1080/15361055.2023.2209131} {\bibfield  {journal} {\bibinfo  {journal} {Fusion Science and Technology}\ }\textbf {\bibinfo {volume} {79}},\ \bibinfo {pages} {1051--1058} (\bibinfo {year} {2023}{\natexlab{a}})},\ \Eprint {http://arxiv.org/abs/https://doi.org/10.1080/15361055.2023.2209131} {https://doi.org/10.1080/15361055.2023.2209131} \BibitemShut {NoStop}%
\bibitem [{\citenamefont {Thompson}\ \emph {et~al.}(2023{\natexlab{b}})\citenamefont {Thompson}, \citenamefont {Simpson}, \citenamefont {Beers}, \citenamefont {Dadras}, \citenamefont {Meier},\ and\ \citenamefont {Stoltz}}]{Thompson_2023_POP}%
  \BibitemOpen
  \bibfield  {author} {\bibinfo {author} {\bibfnamefont {M.~C.}\ \bibnamefont {Thompson}}, \bibinfo {author} {\bibfnamefont {S.~C.}\ \bibnamefont {Simpson}}, \bibinfo {author} {\bibfnamefont {C.~J.}\ \bibnamefont {Beers}}, \bibinfo {author} {\bibfnamefont {J.}~\bibnamefont {Dadras}}, \bibinfo {author} {\bibfnamefont {E.~T.}\ \bibnamefont {Meier}}, \ and\ \bibinfo {author} {\bibfnamefont {P.~H.}\ \bibnamefont {Stoltz}},\ }\bibfield  {title} {\enquote {\bibinfo {title} {{Electrode durability and sheared-flow-stabilized Z-pinch fusion energy}},}\ }\href {\doibase 10.1063/5.0163381} {\bibfield  {journal} {\bibinfo  {journal} {Physics of Plasmas}\ }\textbf {\bibinfo {volume} {30}},\ \bibinfo {pages} {100601} (\bibinfo {year} {2023}{\natexlab{b}})},\ \Eprint {http://arxiv.org/abs/https://doi.org/10.1063/5.0163381} {https://doi.org/10.1063/5.0163381} \BibitemShut {NoStop}%
\bibitem [{\citenamefont {Kimblin}(1974)}]{Kimblin_1974}%
  \BibitemOpen
  \bibfield  {author} {\bibinfo {author} {\bibfnamefont {C.~W.}\ \bibnamefont {Kimblin}},\ }\bibfield  {title} {\enquote {\bibinfo {title} {Cathode spot erosion and ionization phenomena in the transition from vacuum to atmospheric pressure arcs},}\ }\href {\doibase 10.1063/1.1663222} {\bibfield  {journal} {\bibinfo  {journal} {Journal of Applied Physics}\ }\textbf {\bibinfo {volume} {45}},\ \bibinfo {pages} {5235--5244} (\bibinfo {year} {1974})},\ \Eprint {http://arxiv.org/abs/https://doi.org/10.1063/1.1663222} {https://doi.org/10.1063/1.1663222} \BibitemShut {NoStop}%
\bibitem [{\citenamefont {Brown}\ and\ \citenamefont {Shiraishi}(1990)}]{Brown_1990}%
  \BibitemOpen
  \bibfield  {author} {\bibinfo {author} {\bibfnamefont {I.}~\bibnamefont {Brown}}\ and\ \bibinfo {author} {\bibfnamefont {H.}~\bibnamefont {Shiraishi}},\ }\bibfield  {title} {\enquote {\bibinfo {title} {Cathode erosion rates in vacuum-arc discharges},}\ }\href {\doibase 10.1109/27.45521} {\bibfield  {journal} {\bibinfo  {journal} {IEEE Transactions on Plasma Science}\ }\textbf {\bibinfo {volume} {18}},\ \bibinfo {pages} {170--171} (\bibinfo {year} {1990})},\ \Eprint {http://arxiv.org/abs/https://doi.org/10.1109/27.45521} {https://doi.org/10.1109/27.45521} \BibitemShut {NoStop}%
\bibitem [{\citenamefont {Michelis}\ and\ \citenamefont {Mattioli}(1984)}]{DeMichelis_1984}%
  \BibitemOpen
  \bibfield  {author} {\bibinfo {author} {\bibfnamefont {C.~D.}\ \bibnamefont {Michelis}}\ and\ \bibinfo {author} {\bibfnamefont {M.}~\bibnamefont {Mattioli}},\ }\bibfield  {title} {\enquote {\bibinfo {title} {Spectroscopy and impurity behaviour in fusion plasmas},}\ }\href {\doibase 10.1088/0034-4885/47/10/001} {\bibfield  {journal} {\bibinfo  {journal} {Reports on Progress in Physics}\ }\textbf {\bibinfo {volume} {47}},\ \bibinfo {pages} {1233} (\bibinfo {year} {1984})},\ \Eprint {http://arxiv.org/abs/https://dx.doi.org/10.1088/0034-4885/47/10/001} {https://dx.doi.org/10.1088/0034-4885/47/10/001} \BibitemShut {NoStop}%
\bibitem [{\citenamefont {Morgan}\ \emph {et~al.}(1985)\citenamefont {Morgan}, \citenamefont {Behringer}, \citenamefont {Carolan}, \citenamefont {Forrest}, \citenamefont {Peacock},\ and\ \citenamefont {Stamp}}]{Morgan1985}%
  \BibitemOpen
  \bibfield  {author} {\bibinfo {author} {\bibfnamefont {P.~D.}\ \bibnamefont {Morgan}}, \bibinfo {author} {\bibfnamefont {K.~H.}\ \bibnamefont {Behringer}}, \bibinfo {author} {\bibfnamefont {P.~G.}\ \bibnamefont {Carolan}}, \bibinfo {author} {\bibfnamefont {M.~J.}\ \bibnamefont {Forrest}}, \bibinfo {author} {\bibfnamefont {N.~J.}\ \bibnamefont {Peacock}}, \ and\ \bibinfo {author} {\bibfnamefont {M.~F.}\ \bibnamefont {Stamp}},\ }\bibfield  {title} {\enquote {\bibinfo {title} {Spectroscopic measurements on the joint european torus using optical fibers to relay visible radiation},}\ }\href {\doibase 10.1063/1.1138074} {\bibfield  {journal} {\bibinfo  {journal} {Review of Scientific Instruments}\ }\textbf {\bibinfo {volume} {56}},\ \bibinfo {pages} {862--864} (\bibinfo {year} {1985})},\ \Eprint {http://arxiv.org/abs/https://doi.org/10.1063/1.1138074} {https://doi.org/10.1063/1.1138074} \BibitemShut {NoStop}%
\bibitem [{\citenamefont {Behringer}(1986)}]{Behringer_1986RSI}%
  \BibitemOpen
  \bibfield  {author} {\bibinfo {author} {\bibfnamefont {K.}~\bibnamefont {Behringer}},\ }\bibfield  {title} {\enquote {\bibinfo {title} {Spectroscopic diagnostics on jet (invited)},}\ }\href {\doibase 10.1063/1.1138771} {\bibfield  {journal} {\bibinfo  {journal} {Review of Scientific Instruments}\ }\textbf {\bibinfo {volume} {57}},\ \bibinfo {pages} {2000--2005} (\bibinfo {year} {1986})},\ \Eprint {http://arxiv.org/abs/https://doi.org/10.1063/1.1138771} {https://doi.org/10.1063/1.1138771} \BibitemShut {NoStop}%
\bibitem [{\citenamefont {Hintz}\ and\ \citenamefont {Bogen}(1984)}]{HINTZ1984229}%
  \BibitemOpen
  \bibfield  {author} {\bibinfo {author} {\bibfnamefont {E.}~\bibnamefont {Hintz}}\ and\ \bibinfo {author} {\bibfnamefont {P.}~\bibnamefont {Bogen}},\ }\bibfield  {title} {\enquote {\bibinfo {title} {Plasma edge diagnostics by optical methods},}\ }\href {\doibase https://doi.org/10.1016/0022-3115(84)90357-X} {\bibfield  {journal} {\bibinfo  {journal} {Journal of Nuclear Materials}\ }\textbf {\bibinfo {volume} {128-129}},\ \bibinfo {pages} {229--239} (\bibinfo {year} {1984})},\ \Eprint {http://arxiv.org/abs/https://doi.org/10.1016/0022-3115(84)90357-X} {https://doi.org/10.1016/0022-3115(84)90357-X} \BibitemShut {NoStop}%
\bibitem [{\citenamefont {Fussmann}\ \emph {et~al.}(1984)\citenamefont {Fussmann}, \citenamefont {Ditte}, \citenamefont {Eckstein}, \citenamefont {Grave}, \citenamefont {Keilhacker}, \citenamefont {McCormick}, \citenamefont {Murmann}, \citenamefont {Röhr}, \citenamefont {Elshaer}, \citenamefont {Steuer}, \citenamefont {Szymanski}, \citenamefont {Wagner}, \citenamefont {Becker}, \citenamefont {Bernhardi}, \citenamefont {Eberhagen}, \citenamefont {Gehre}, \citenamefont {Gernhardt}, \citenamefont {Gierke}, \citenamefont {Glock}, \citenamefont {Gruber}, \citenamefont {Haas}, \citenamefont {Hesse}, \citenamefont {Janeschitz}, \citenamefont {Karger}, \citenamefont {Kissel}, \citenamefont {Klüber}, \citenamefont {Kornherr}, \citenamefont {Lisitano}, \citenamefont {Mayer}, \citenamefont {Meisel}, \citenamefont {Müller}, \citenamefont {Poschenrieder}, \citenamefont {Ryter}, \citenamefont {Rapp}, \citenamefont {Schneider}, \citenamefont {Siller}, \citenamefont {Smeulders}, \citenamefont {Söldner}, \citenamefont
  {Speth}, \citenamefont {Stäbler},\ and\ \citenamefont {Vollmer}}]{FUSSMANN1984350}%
  \BibitemOpen
  \bibfield  {author} {\bibinfo {author} {\bibfnamefont {G.}~\bibnamefont {Fussmann}}, \bibinfo {author} {\bibfnamefont {U.}~\bibnamefont {Ditte}}, \bibinfo {author} {\bibfnamefont {W.}~\bibnamefont {Eckstein}}, \bibinfo {author} {\bibfnamefont {T.}~\bibnamefont {Grave}}, \bibinfo {author} {\bibfnamefont {M.}~\bibnamefont {Keilhacker}}, \bibinfo {author} {\bibfnamefont {K.}~\bibnamefont {McCormick}}, \bibinfo {author} {\bibfnamefont {H.}~\bibnamefont {Murmann}}, \bibinfo {author} {\bibfnamefont {H.}~\bibnamefont {Röhr}}, \bibinfo {author} {\bibfnamefont {M.}~\bibnamefont {Elshaer}}, \bibinfo {author} {\bibfnamefont {K.-H.}\ \bibnamefont {Steuer}}, \bibinfo {author} {\bibfnamefont {Z.}~\bibnamefont {Szymanski}}, \bibinfo {author} {\bibfnamefont {F.}~\bibnamefont {Wagner}}, \bibinfo {author} {\bibfnamefont {G.}~\bibnamefont {Becker}}, \bibinfo {author} {\bibfnamefont {K.}~\bibnamefont {Bernhardi}}, \bibinfo {author} {\bibfnamefont {A.}~\bibnamefont {Eberhagen}}, \bibinfo {author} {\bibfnamefont
  {O.}~\bibnamefont {Gehre}}, \bibinfo {author} {\bibfnamefont {J.}~\bibnamefont {Gernhardt}}, \bibinfo {author} {\bibfnamefont {G.}~\bibnamefont {Gierke}}, \bibinfo {author} {\bibfnamefont {E.}~\bibnamefont {Glock}}, \bibinfo {author} {\bibfnamefont {O.}~\bibnamefont {Gruber}}, \bibinfo {author} {\bibfnamefont {G.}~\bibnamefont {Haas}}, \bibinfo {author} {\bibfnamefont {M.}~\bibnamefont {Hesse}}, \bibinfo {author} {\bibfnamefont {G.}~\bibnamefont {Janeschitz}}, \bibinfo {author} {\bibfnamefont {F.}~\bibnamefont {Karger}}, \bibinfo {author} {\bibfnamefont {S.}~\bibnamefont {Kissel}}, \bibinfo {author} {\bibfnamefont {O.}~\bibnamefont {Klüber}}, \bibinfo {author} {\bibfnamefont {M.}~\bibnamefont {Kornherr}}, \bibinfo {author} {\bibfnamefont {G.}~\bibnamefont {Lisitano}}, \bibinfo {author} {\bibfnamefont {H.}~\bibnamefont {Mayer}}, \bibinfo {author} {\bibfnamefont {D.}~\bibnamefont {Meisel}}, \bibinfo {author} {\bibfnamefont {E.}~\bibnamefont {Müller}}, \bibinfo {author} {\bibfnamefont {W.}~\bibnamefont
  {Poschenrieder}}, \bibinfo {author} {\bibfnamefont {F.}~\bibnamefont {Ryter}}, \bibinfo {author} {\bibfnamefont {H.}~\bibnamefont {Rapp}}, \bibinfo {author} {\bibfnamefont {F.}~\bibnamefont {Schneider}}, \bibinfo {author} {\bibfnamefont {G.}~\bibnamefont {Siller}}, \bibinfo {author} {\bibfnamefont {P.}~\bibnamefont {Smeulders}}, \bibinfo {author} {\bibfnamefont {F.}~\bibnamefont {Söldner}}, \bibinfo {author} {\bibfnamefont {E.}~\bibnamefont {Speth}}, \bibinfo {author} {\bibfnamefont {A.}~\bibnamefont {Stäbler}}, \ and\ \bibinfo {author} {\bibfnamefont {O.}~\bibnamefont {Vollmer}},\ }\bibfield  {title} {\enquote {\bibinfo {title} {Divertor parameters and divertor operation in asdex},}\ }\href {\doibase https://doi.org/10.1016/0022-3115(84)90377-5} {\bibfield  {journal} {\bibinfo  {journal} {Journal of Nuclear Materials}\ }\textbf {\bibinfo {volume} {128-129}},\ \bibinfo {pages} {350--358} (\bibinfo {year} {1984})},\ \Eprint {http://arxiv.org/abs/https://doi.org/10.1016/0022-3115(84)90377-5}
  {https://doi.org/10.1016/0022-3115(84)90377-5} \BibitemShut {NoStop}%
\bibitem [{\citenamefont {Rudakov}\ \emph {et~al.}(2017)\citenamefont {Rudakov}, \citenamefont {Abrams}, \citenamefont {Ding}, \citenamefont {Guo}, \citenamefont {Stangeby}, \citenamefont {Wampler}, \citenamefont {Boedo}, \citenamefont {Briesemeister}, \citenamefont {Brooks}, \citenamefont {Buchenauer}, \citenamefont {Bykov}, \citenamefont {Chrobak}, \citenamefont {Doerner}, \citenamefont {Donovan}, \citenamefont {Elder}, \citenamefont {Fenstermacher}, \citenamefont {Guterl}, \citenamefont {Hinson}, \citenamefont {Hollmann}, \citenamefont {Lasnier}, \citenamefont {Leonard}, \citenamefont {McLean}, \citenamefont {Moyer}, \citenamefont {Nygren}, \citenamefont {Thomas}, \citenamefont {Unterberg}, \citenamefont {Watkins},\ and\ \citenamefont {Wong}}]{RUDAKOV2017196}%
  \BibitemOpen
  \bibfield  {author} {\bibinfo {author} {\bibfnamefont {D.}~\bibnamefont {Rudakov}}, \bibinfo {author} {\bibfnamefont {T.}~\bibnamefont {Abrams}}, \bibinfo {author} {\bibfnamefont {R.}~\bibnamefont {Ding}}, \bibinfo {author} {\bibfnamefont {H.}~\bibnamefont {Guo}}, \bibinfo {author} {\bibfnamefont {P.}~\bibnamefont {Stangeby}}, \bibinfo {author} {\bibfnamefont {W.}~\bibnamefont {Wampler}}, \bibinfo {author} {\bibfnamefont {J.}~\bibnamefont {Boedo}}, \bibinfo {author} {\bibfnamefont {A.}~\bibnamefont {Briesemeister}}, \bibinfo {author} {\bibfnamefont {J.}~\bibnamefont {Brooks}}, \bibinfo {author} {\bibfnamefont {D.}~\bibnamefont {Buchenauer}}, \bibinfo {author} {\bibfnamefont {I.}~\bibnamefont {Bykov}}, \bibinfo {author} {\bibfnamefont {C.}~\bibnamefont {Chrobak}}, \bibinfo {author} {\bibfnamefont {R.}~\bibnamefont {Doerner}}, \bibinfo {author} {\bibfnamefont {D.}~\bibnamefont {Donovan}}, \bibinfo {author} {\bibfnamefont {J.}~\bibnamefont {Elder}}, \bibinfo {author} {\bibfnamefont {M.}~\bibnamefont
  {Fenstermacher}}, \bibinfo {author} {\bibfnamefont {J.}~\bibnamefont {Guterl}}, \bibinfo {author} {\bibfnamefont {E.}~\bibnamefont {Hinson}}, \bibinfo {author} {\bibfnamefont {E.}~\bibnamefont {Hollmann}}, \bibinfo {author} {\bibfnamefont {C.}~\bibnamefont {Lasnier}}, \bibinfo {author} {\bibfnamefont {A.}~\bibnamefont {Leonard}}, \bibinfo {author} {\bibfnamefont {A.}~\bibnamefont {McLean}}, \bibinfo {author} {\bibfnamefont {R.}~\bibnamefont {Moyer}}, \bibinfo {author} {\bibfnamefont {R.}~\bibnamefont {Nygren}}, \bibinfo {author} {\bibfnamefont {D.}~\bibnamefont {Thomas}}, \bibinfo {author} {\bibfnamefont {E.}~\bibnamefont {Unterberg}}, \bibinfo {author} {\bibfnamefont {J.}~\bibnamefont {Watkins}}, \ and\ \bibinfo {author} {\bibfnamefont {C.}~\bibnamefont {Wong}},\ }\bibfield  {title} {\enquote {\bibinfo {title} {Dimes pmi research at diii-d in support of iter and beyond},}\ }\href {\doibase https://doi.org/10.1016/j.fusengdes.2017.03.007} {\bibfield  {journal} {\bibinfo  {journal} {Fusion Engineering and
  Design}\ }\textbf {\bibinfo {volume} {124}},\ \bibinfo {pages} {196--201} (\bibinfo {year} {2017})},\ \bibinfo {note} {proceedings of the 29th Symposium on Fusion Technology (SOFT-29) Prague, Czech Republic, September 5-9, 2016},\ \Eprint {http://arxiv.org/abs/https://doi.org/10.1016/j.fusengdes.2017.03.007} {https://doi.org/10.1016/j.fusengdes.2017.03.007} \BibitemShut {NoStop}%
\bibitem [{\citenamefont {Abrams}\ \emph {et~al.}(2021)\citenamefont {Abrams}, \citenamefont {Bringuier}, \citenamefont {Thomas}, \citenamefont {Sinclair}, \citenamefont {Gonderman}, \citenamefont {Holland}, \citenamefont {Rudakov}, \citenamefont {Wilcox}, \citenamefont {Unterberg},\ and\ \citenamefont {Scotti}}]{Abrams_2021}%
  \BibitemOpen
  \bibfield  {author} {\bibinfo {author} {\bibfnamefont {T.}~\bibnamefont {Abrams}}, \bibinfo {author} {\bibfnamefont {S.}~\bibnamefont {Bringuier}}, \bibinfo {author} {\bibfnamefont {D.}~\bibnamefont {Thomas}}, \bibinfo {author} {\bibfnamefont {G.}~\bibnamefont {Sinclair}}, \bibinfo {author} {\bibfnamefont {S.}~\bibnamefont {Gonderman}}, \bibinfo {author} {\bibfnamefont {L.}~\bibnamefont {Holland}}, \bibinfo {author} {\bibfnamefont {D.}~\bibnamefont {Rudakov}}, \bibinfo {author} {\bibfnamefont {R.}~\bibnamefont {Wilcox}}, \bibinfo {author} {\bibfnamefont {E.}~\bibnamefont {Unterberg}}, \ and\ \bibinfo {author} {\bibfnamefont {F.}~\bibnamefont {Scotti}},\ }\bibfield  {title} {\enquote {\bibinfo {title} {Evaluation of silicon carbide as a divertor armor material in diii-d h-mode discharges},}\ }\href {\doibase 10.1088/1741-4326/abecee} {\bibfield  {journal} {\bibinfo  {journal} {Nuclear Fusion}\ }\textbf {\bibinfo {volume} {61}},\ \bibinfo {pages} {066005} (\bibinfo {year} {2021})},\ \Eprint
  {http://arxiv.org/abs/https://dx.doi.org/10.1088/1741-4326/abecee} {https://dx.doi.org/10.1088/1741-4326/abecee} \BibitemShut {NoStop}%
\bibitem [{\citenamefont {James}\ \emph {et~al.}(2013)\citenamefont {James}, \citenamefont {Brunner}, \citenamefont {Labombard}, \citenamefont {Lau}, \citenamefont {Lipschultz}, \citenamefont {Miller}, \citenamefont {Reinke}, \citenamefont {Terry}, \citenamefont {Theiler}, \citenamefont {Wallace}, \citenamefont {Whyte}, \citenamefont {Wukitch},\ and\ \citenamefont {Soukhanovskii}}]{James_2013}%
  \BibitemOpen
  \bibfield  {author} {\bibinfo {author} {\bibfnamefont {A.~N.}\ \bibnamefont {James}}, \bibinfo {author} {\bibfnamefont {D.}~\bibnamefont {Brunner}}, \bibinfo {author} {\bibfnamefont {B.}~\bibnamefont {Labombard}}, \bibinfo {author} {\bibfnamefont {C.}~\bibnamefont {Lau}}, \bibinfo {author} {\bibfnamefont {B.}~\bibnamefont {Lipschultz}}, \bibinfo {author} {\bibfnamefont {D.}~\bibnamefont {Miller}}, \bibinfo {author} {\bibfnamefont {M.~L.}\ \bibnamefont {Reinke}}, \bibinfo {author} {\bibfnamefont {J.~L.}\ \bibnamefont {Terry}}, \bibinfo {author} {\bibfnamefont {C.}~\bibnamefont {Theiler}}, \bibinfo {author} {\bibfnamefont {G.~M.}\ \bibnamefont {Wallace}}, \bibinfo {author} {\bibfnamefont {D.~G.}\ \bibnamefont {Whyte}}, \bibinfo {author} {\bibfnamefont {S.}~\bibnamefont {Wukitch}}, \ and\ \bibinfo {author} {\bibfnamefont {V.}~\bibnamefont {Soukhanovskii}},\ }\bibfield  {title} {\enquote {\bibinfo {title} {Imaging of molybdenum erosion and thermography at visible wavelengths in alcator c-mod icrh and lhcd
  discharges},}\ }\href {\doibase 10.1088/0741-3335/55/12/125010} {\bibfield  {journal} {\bibinfo  {journal} {Plasma Physics and Controlled Fusion}\ }\textbf {\bibinfo {volume} {55}},\ \bibinfo {pages} {125010} (\bibinfo {year} {2013})},\ \Eprint {http://arxiv.org/abs/https://dx.doi.org/10.1088/0741-3335/55/12/125010} {https://dx.doi.org/10.1088/0741-3335/55/12/125010} \BibitemShut {NoStop}%
\bibitem [{\citenamefont {Wong}\ \emph {et~al.}(2007)\citenamefont {Wong}, \citenamefont {Rudakov}, \citenamefont {Allain}, \citenamefont {Bastasz}, \citenamefont {Brooks}, \citenamefont {Brooks}, \citenamefont {Doerner}, \citenamefont {Evans}, \citenamefont {Hassanein}, \citenamefont {Jacob}, \citenamefont {Krieger}, \citenamefont {Litnovsky}, \citenamefont {McLean}, \citenamefont {Philipps}, \citenamefont {Pigarov}, \citenamefont {Wampler}, \citenamefont {Watkins}, \citenamefont {West}, \citenamefont {Whaley},\ and\ \citenamefont {Wienhold}}]{WONG2007276}%
  \BibitemOpen
  \bibfield  {author} {\bibinfo {author} {\bibfnamefont {C.}~\bibnamefont {Wong}}, \bibinfo {author} {\bibfnamefont {D.}~\bibnamefont {Rudakov}}, \bibinfo {author} {\bibfnamefont {J.}~\bibnamefont {Allain}}, \bibinfo {author} {\bibfnamefont {R.}~\bibnamefont {Bastasz}}, \bibinfo {author} {\bibfnamefont {N.}~\bibnamefont {Brooks}}, \bibinfo {author} {\bibfnamefont {J.}~\bibnamefont {Brooks}}, \bibinfo {author} {\bibfnamefont {R.}~\bibnamefont {Doerner}}, \bibinfo {author} {\bibfnamefont {T.}~\bibnamefont {Evans}}, \bibinfo {author} {\bibfnamefont {A.}~\bibnamefont {Hassanein}}, \bibinfo {author} {\bibfnamefont {W.}~\bibnamefont {Jacob}}, \bibinfo {author} {\bibfnamefont {K.}~\bibnamefont {Krieger}}, \bibinfo {author} {\bibfnamefont {A.}~\bibnamefont {Litnovsky}}, \bibinfo {author} {\bibfnamefont {A.}~\bibnamefont {McLean}}, \bibinfo {author} {\bibfnamefont {V.}~\bibnamefont {Philipps}}, \bibinfo {author} {\bibfnamefont {A.}~\bibnamefont {Pigarov}}, \bibinfo {author} {\bibfnamefont {W.}~\bibnamefont {Wampler}},
  \bibinfo {author} {\bibfnamefont {J.}~\bibnamefont {Watkins}}, \bibinfo {author} {\bibfnamefont {W.}~\bibnamefont {West}}, \bibinfo {author} {\bibfnamefont {J.}~\bibnamefont {Whaley}}, \ and\ \bibinfo {author} {\bibfnamefont {P.}~\bibnamefont {Wienhold}},\ }\bibfield  {title} {\enquote {\bibinfo {title} {Divertor and midplane materials evaluation system in diii-d},}\ }\href {\doibase https://doi.org/10.1016/j.jnucmat.2007.01.121} {\bibfield  {journal} {\bibinfo  {journal} {Journal of Nuclear Materials}\ }\textbf {\bibinfo {volume} {363-365}},\ \bibinfo {pages} {276--281} (\bibinfo {year} {2007})},\ \Eprint {http://arxiv.org/abs/https://doi.org/10.1016/j.jnucmat.2007.01.121} {https://doi.org/10.1016/j.jnucmat.2007.01.121} \BibitemShut {NoStop}%
\bibitem [{\citenamefont {Naujoks}\ \emph {et~al.}(1996)\citenamefont {Naujoks}, \citenamefont {Asmussen}, \citenamefont {Bessenrodt-Weberpals}, \citenamefont {Deschka}, \citenamefont {Dux}, \citenamefont {Engelhardt}, \citenamefont {Field}, \citenamefont {Fussmann}, \citenamefont {Fuchs}, \citenamefont {Garcia-Rosales}, \citenamefont {Hirsch}, \citenamefont {Ignacz}, \citenamefont {Lieder}, \citenamefont {Mast}, \citenamefont {Neu}, \citenamefont {Radtke}, \citenamefont {Roth},\ and\ \citenamefont {Wenzel}}]{Naujoks_1996}%
  \BibitemOpen
  \bibfield  {author} {\bibinfo {author} {\bibfnamefont {D.}~\bibnamefont {Naujoks}}, \bibinfo {author} {\bibfnamefont {K.}~\bibnamefont {Asmussen}}, \bibinfo {author} {\bibfnamefont {M.}~\bibnamefont {Bessenrodt-Weberpals}}, \bibinfo {author} {\bibfnamefont {S.}~\bibnamefont {Deschka}}, \bibinfo {author} {\bibfnamefont {R.}~\bibnamefont {Dux}}, \bibinfo {author} {\bibfnamefont {W.}~\bibnamefont {Engelhardt}}, \bibinfo {author} {\bibfnamefont {A.}~\bibnamefont {Field}}, \bibinfo {author} {\bibfnamefont {G.}~\bibnamefont {Fussmann}}, \bibinfo {author} {\bibfnamefont {J.}~\bibnamefont {Fuchs}}, \bibinfo {author} {\bibfnamefont {C.}~\bibnamefont {Garcia-Rosales}}, \bibinfo {author} {\bibfnamefont {S.}~\bibnamefont {Hirsch}}, \bibinfo {author} {\bibfnamefont {P.}~\bibnamefont {Ignacz}}, \bibinfo {author} {\bibfnamefont {G.}~\bibnamefont {Lieder}}, \bibinfo {author} {\bibfnamefont {K.}~\bibnamefont {Mast}}, \bibinfo {author} {\bibfnamefont {R.}~\bibnamefont {Neu}}, \bibinfo {author} {\bibfnamefont
  {R.}~\bibnamefont {Radtke}}, \bibinfo {author} {\bibfnamefont {J.}~\bibnamefont {Roth}}, \ and\ \bibinfo {author} {\bibfnamefont {U.}~\bibnamefont {Wenzel}},\ }\bibfield  {title} {\enquote {\bibinfo {title} {Tungsten as target material in fusion devices},}\ }\href {\doibase 10.1088/0029-5515/36/6/I01} {\bibfield  {journal} {\bibinfo  {journal} {Nuclear Fusion}\ }\textbf {\bibinfo {volume} {36}},\ \bibinfo {pages} {671} (\bibinfo {year} {1996})},\ \Eprint {http://arxiv.org/abs/https://dx.doi.org/10.1088/0029-5515/36/6/I01} {https://dx.doi.org/10.1088/0029-5515/36/6/I01} \BibitemShut {NoStop}%
\bibitem [{\citenamefont {Roth}\ \emph {et~al.}(1995)\citenamefont {Roth}, \citenamefont {Naujoks}, \citenamefont {Krieger}, \citenamefont {Field}, \citenamefont {Lieder},\ and\ \citenamefont {Hirsch}}]{ROTH1995231}%
  \BibitemOpen
  \bibfield  {author} {\bibinfo {author} {\bibfnamefont {J.}~\bibnamefont {Roth}}, \bibinfo {author} {\bibfnamefont {D.}~\bibnamefont {Naujoks}}, \bibinfo {author} {\bibfnamefont {K.}~\bibnamefont {Krieger}}, \bibinfo {author} {\bibfnamefont {A.}~\bibnamefont {Field}}, \bibinfo {author} {\bibfnamefont {G.}~\bibnamefont {Lieder}}, \ and\ \bibinfo {author} {\bibfnamefont {S.}~\bibnamefont {Hirsch}},\ }\bibfield  {title} {\enquote {\bibinfo {title} {Experimental investigations of high-z materials in the asdex-upgrade divertor},}\ }\href {\doibase https://doi.org/10.1016/0022-3115(94)00418-8} {\bibfield  {journal} {\bibinfo  {journal} {Journal of Nuclear Materials}\ }\textbf {\bibinfo {volume} {220-222}},\ \bibinfo {pages} {231--234} (\bibinfo {year} {1995})},\ \bibinfo {note} {plasma-Surface Interactions in Controlled Fusion Devices},\ \Eprint {http://arxiv.org/abs/https://doi.org/10.1016/0022-3115(94)00418-8} {https://doi.org/10.1016/0022-3115(94)00418-8} \BibitemShut {NoStop}%
\bibitem [{\citenamefont {Behringer}\ \emph {et~al.}(1989)\citenamefont {Behringer}, \citenamefont {Summers}, \citenamefont {Denne}, \citenamefont {Forrest},\ and\ \citenamefont {Stamp}}]{Behringer_1989}%
  \BibitemOpen
  \bibfield  {author} {\bibinfo {author} {\bibfnamefont {K.}~\bibnamefont {Behringer}}, \bibinfo {author} {\bibfnamefont {H.~P.}\ \bibnamefont {Summers}}, \bibinfo {author} {\bibfnamefont {B.}~\bibnamefont {Denne}}, \bibinfo {author} {\bibfnamefont {M.}~\bibnamefont {Forrest}}, \ and\ \bibinfo {author} {\bibfnamefont {M.}~\bibnamefont {Stamp}},\ }\bibfield  {title} {\enquote {\bibinfo {title} {Spectroscopic determination of impurity influx from localized surfaces},}\ }\href {\doibase 10.1088/0741-3335/31/14/001} {\bibfield  {journal} {\bibinfo  {journal} {Plasma Physics and Controlled Fusion}\ }\textbf {\bibinfo {volume} {31}},\ \bibinfo {pages} {2059--2099} (\bibinfo {year} {1989})},\ \Eprint {http://arxiv.org/abs/https://doi.org/10.1088/0741-3335/31/14/001} {https://doi.org/10.1088/0741-3335/31/14/001} \BibitemShut {NoStop}%
\bibitem [{\citenamefont {Den~Hartog}\ and\ \citenamefont {Golingo}(2001)}]{DenHartog_2001_RSI}%
  \BibitemOpen
  \bibfield  {author} {\bibinfo {author} {\bibfnamefont {D.~J.}\ \bibnamefont {Den~Hartog}}\ and\ \bibinfo {author} {\bibfnamefont {R.~P.}\ \bibnamefont {Golingo}},\ }\bibfield  {title} {\enquote {\bibinfo {title} {Telecentric viewing system for light collection from a z-pinch plasma},}\ }\href {\doibase 10.1063/1.1353188} {\bibfield  {journal} {\bibinfo  {journal} {Review of Scientific Instruments}\ }\textbf {\bibinfo {volume} {72}},\ \bibinfo {pages} {2224--2225} (\bibinfo {year} {2001})},\ \Eprint {http://arxiv.org/abs/https://doi.org/10.1063/1.1353188} {https://doi.org/10.1063/1.1353188} \BibitemShut {NoStop}%
\bibitem [{\citenamefont {Forbes}\ and\ \citenamefont {Shumlak}(2020)}]{Forbes_2020_RSI}%
  \BibitemOpen
  \bibfield  {author} {\bibinfo {author} {\bibfnamefont {E.~G.}\ \bibnamefont {Forbes}}\ and\ \bibinfo {author} {\bibfnamefont {U.}~\bibnamefont {Shumlak}},\ }\bibfield  {title} {\enquote {\bibinfo {title} {Spatio-temporal ion temperature and velocity measurements in a z pinch using fast-framing spectroscopy},}\ }\href {\doibase 10.1063/5.0012255} {\bibfield  {journal} {\bibinfo  {journal} {Review of Scientific Instruments}\ }\textbf {\bibinfo {volume} {91}},\ \bibinfo {pages} {083104} (\bibinfo {year} {2020})},\ \Eprint {http://arxiv.org/abs/https://doi.org/10.1063/5.0012255} {https://doi.org/10.1063/5.0012255} \BibitemShut {NoStop}%
\bibitem [{\citenamefont {Summers}\ and\ \citenamefont {O’Mullane}(2011)}]{Summers_2011_ADAS}%
  \BibitemOpen
  \bibfield  {author} {\bibinfo {author} {\bibfnamefont {H.~P.}\ \bibnamefont {Summers}}\ and\ \bibinfo {author} {\bibfnamefont {M.~G.}\ \bibnamefont {O’Mullane}},\ }\bibfield  {title} {\enquote {\bibinfo {title} {{Atomic Data and Modelling for Fusion: the ADAS Project}},}\ }\href {\doibase 10.1063/1.3585817} {\bibfield  {journal} {\bibinfo  {journal} {AIP Conference Proceedings}\ }\textbf {\bibinfo {volume} {1344}},\ \bibinfo {pages} {179--187} (\bibinfo {year} {2011})},\ \Eprint {http://arxiv.org/abs/https://doi.org/10.1063/1.3585817} {https://doi.org/10.1063/1.3585817} \BibitemShut {NoStop}%
\bibitem [{\citenamefont {Shumlak}\ \emph {et~al.}(2001)\citenamefont {Shumlak}, \citenamefont {Golingo}, \citenamefont {Nelson},\ and\ \citenamefont {Den~Hartog}}]{Shumlak_2001_PRL}%
  \BibitemOpen
  \bibfield  {author} {\bibinfo {author} {\bibfnamefont {U.}~\bibnamefont {Shumlak}}, \bibinfo {author} {\bibfnamefont {R.~P.}\ \bibnamefont {Golingo}}, \bibinfo {author} {\bibfnamefont {B.~A.}\ \bibnamefont {Nelson}}, \ and\ \bibinfo {author} {\bibfnamefont {D.~J.}\ \bibnamefont {Den~Hartog}},\ }\bibfield  {title} {\enquote {\bibinfo {title} {Evidence of stabilization in the $\mathit{Z}$-pinch},}\ }\href {\doibase 10.1103/PhysRevLett.87.205005} {\bibfield  {journal} {\bibinfo  {journal} {Phys. Rev. Lett.}\ }\textbf {\bibinfo {volume} {87}},\ \bibinfo {pages} {205005} (\bibinfo {year} {2001})},\ \Eprint {http://arxiv.org/abs/https://doi.org/10.1103/PhysRevLett.87.205005} {https://doi.org/10.1103/PhysRevLett.87.205005} \BibitemShut {NoStop}%
\bibitem [{\citenamefont {Golingo}, \citenamefont {Shumlak},\ and\ \citenamefont {Nelson}(2005)}]{Golingo_2005}%
  \BibitemOpen
  \bibfield  {author} {\bibinfo {author} {\bibfnamefont {R.~P.}\ \bibnamefont {Golingo}}, \bibinfo {author} {\bibfnamefont {U.}~\bibnamefont {Shumlak}}, \ and\ \bibinfo {author} {\bibfnamefont {B.~A.}\ \bibnamefont {Nelson}},\ }\bibfield  {title} {\enquote {\bibinfo {title} {Formation of a sheared flow z pinch},}\ }\href {\doibase 10.1063/1.1928249} {\bibfield  {journal} {\bibinfo  {journal} {Physics of Plasmas}\ }\textbf {\bibinfo {volume} {12}},\ \bibinfo {pages} {062505} (\bibinfo {year} {2005})},\ \Eprint {http://arxiv.org/abs/https://doi.org/10.1063/1.1928249} {https://doi.org/10.1063/1.1928249} \BibitemShut {NoStop}%
\bibitem [{\citenamefont {Shumlak}\ \emph {et~al.}(2009)\citenamefont {Shumlak}, \citenamefont {Adams}, \citenamefont {Blakely}, \citenamefont {Chan}, \citenamefont {Golingo}, \citenamefont {Knecht}, \citenamefont {Nelson}, \citenamefont {Oberto}, \citenamefont {Sybouts},\ and\ \citenamefont {Vogman}}]{Shumlak_2009}%
  \BibitemOpen
  \bibfield  {author} {\bibinfo {author} {\bibfnamefont {U.}~\bibnamefont {Shumlak}}, \bibinfo {author} {\bibfnamefont {C.}~\bibnamefont {Adams}}, \bibinfo {author} {\bibfnamefont {J.}~\bibnamefont {Blakely}}, \bibinfo {author} {\bibfnamefont {B.-J.}\ \bibnamefont {Chan}}, \bibinfo {author} {\bibfnamefont {R.}~\bibnamefont {Golingo}}, \bibinfo {author} {\bibfnamefont {S.}~\bibnamefont {Knecht}}, \bibinfo {author} {\bibfnamefont {B.}~\bibnamefont {Nelson}}, \bibinfo {author} {\bibfnamefont {R.}~\bibnamefont {Oberto}}, \bibinfo {author} {\bibfnamefont {M.}~\bibnamefont {Sybouts}}, \ and\ \bibinfo {author} {\bibfnamefont {G.}~\bibnamefont {Vogman}},\ }\bibfield  {title} {\enquote {\bibinfo {title} {Equilibrium, flow shear and stability measurements in the z-pinch},}\ }\href {\doibase 10.1088/0029-5515/49/7/075039} {\bibfield  {journal} {\bibinfo  {journal} {Nuclear Fusion}\ }\textbf {\bibinfo {volume} {49}},\ \bibinfo {pages} {075039} (\bibinfo {year} {2009})},\ \Eprint
  {http://arxiv.org/abs/https://dx.doi.org/10.1088/0029-5515/49/7/075039} {https://dx.doi.org/10.1088/0029-5515/49/7/075039} \BibitemShut {NoStop}%
\bibitem [{\citenamefont {Shumlak}\ and\ \citenamefont {Hartman}(1995)}]{Shumlak1995}%
  \BibitemOpen
  \bibfield  {author} {\bibinfo {author} {\bibfnamefont {U.}~\bibnamefont {Shumlak}}\ and\ \bibinfo {author} {\bibfnamefont {C.~W.}\ \bibnamefont {Hartman}},\ }\bibfield  {title} {\enquote {\bibinfo {title} {Sheared flow stabilization of the $\mathit{m}\phantom{\rule{0ex}{0ex}}=\phantom{\rule{0ex}{0ex}}1$ kink mode in $\mathit{Z}$ pinches},}\ }\href {\doibase 10.1103/PhysRevLett.75.3285} {\bibfield  {journal} {\bibinfo  {journal} {Phys. Rev. Lett.}\ }\textbf {\bibinfo {volume} {75}},\ \bibinfo {pages} {3285--3288} (\bibinfo {year} {1995})},\ \Eprint {http://arxiv.org/abs/https://doi.org/10.1103/PhysRevLett.75.3285} {https://doi.org/10.1103/PhysRevLett.75.3285} \BibitemShut {NoStop}%
\bibitem [{\citenamefont {Levitt}\ \emph {et~al.}(2023)\citenamefont {Levitt}, \citenamefont {Meier}, \citenamefont {Umstattd}, \citenamefont {Barhydt}, \citenamefont {Datta}, \citenamefont {Liekhus-Schmaltz}, \citenamefont {Sutherland},\ and\ \citenamefont {Nelson}}]{Levitt_2023}%
  \BibitemOpen
  \bibfield  {author} {\bibinfo {author} {\bibfnamefont {B.}~\bibnamefont {Levitt}}, \bibinfo {author} {\bibfnamefont {E.~T.}\ \bibnamefont {Meier}}, \bibinfo {author} {\bibfnamefont {R.}~\bibnamefont {Umstattd}}, \bibinfo {author} {\bibfnamefont {J.~R.}\ \bibnamefont {Barhydt}}, \bibinfo {author} {\bibfnamefont {I.~A.~M.}\ \bibnamefont {Datta}}, \bibinfo {author} {\bibfnamefont {C.}~\bibnamefont {Liekhus-Schmaltz}}, \bibinfo {author} {\bibfnamefont {D.~A.}\ \bibnamefont {Sutherland}}, \ and\ \bibinfo {author} {\bibfnamefont {B.~A.}\ \bibnamefont {Nelson}},\ }\bibfield  {title} {\enquote {\bibinfo {title} {{The Zap Energy approach to commercial fusion}},}\ }\href {\doibase 10.1063/5.0163361} {\bibfield  {journal} {\bibinfo  {journal} {Physics of Plasmas}\ }\textbf {\bibinfo {volume} {30}},\ \bibinfo {pages} {090603} (\bibinfo {year} {2023})},\ \Eprint {http://arxiv.org/abs/https://doi.org/10.1063/5.0163361} {https://doi.org/10.1063/5.0163361} \BibitemShut {NoStop}%
\bibitem [{\citenamefont {Bennett}(1934)}]{Bennett1934}%
  \BibitemOpen
  \bibfield  {author} {\bibinfo {author} {\bibfnamefont {W.~H.}\ \bibnamefont {Bennett}},\ }\bibfield  {title} {\enquote {\bibinfo {title} {Magnetically self-focussing streams},}\ }\href {\doibase 10.1103/PhysRev.45.890} {\bibfield  {journal} {\bibinfo  {journal} {Phys. Rev.}\ }\textbf {\bibinfo {volume} {45}},\ \bibinfo {pages} {890--897} (\bibinfo {year} {1934})},\ \Eprint {http://arxiv.org/abs/https://doi.org/10.1103/PhysRev.45.890} {https://doi.org/10.1103/PhysRev.45.890} \BibitemShut {NoStop}%
\bibitem [{\citenamefont {Newcomb}(1960)}]{NEWCOMB1960232}%
  \BibitemOpen
  \bibfield  {author} {\bibinfo {author} {\bibfnamefont {W.~A.}\ \bibnamefont {Newcomb}},\ }\bibfield  {title} {\enquote {\bibinfo {title} {Hydromagnetic stability of a diffuse linear pinch},}\ }\href {\doibase https://doi.org/10.1016/0003-4916(60)90023-3} {\bibfield  {journal} {\bibinfo  {journal} {Annals of Physics}\ }\textbf {\bibinfo {volume} {10}},\ \bibinfo {pages} {232--267} (\bibinfo {year} {1960})},\ \Eprint {http://arxiv.org/abs/https://doi.org/10.1016/0003-4916(60)90023-3} {https://doi.org/10.1016/0003-4916(60)90023-3} \BibitemShut {NoStop}%
\bibitem [{\citenamefont {Shumlak}\ \emph {et~al.}(2012)\citenamefont {Shumlak}, \citenamefont {Chadney}, \citenamefont {Golingo}, \citenamefont {Hartog}, \citenamefont {Hughes}, \citenamefont {Knecht}, \citenamefont {Lowrie}, \citenamefont {Lukin}, \citenamefont {Nelson}, \citenamefont {Oberto}, \citenamefont {Rohrbach}, \citenamefont {Ross},\ and\ \citenamefont {Vogman}}]{Shumlak2012_FST}%
  \BibitemOpen
  \bibfield  {author} {\bibinfo {author} {\bibfnamefont {U.}~\bibnamefont {Shumlak}}, \bibinfo {author} {\bibfnamefont {J.}~\bibnamefont {Chadney}}, \bibinfo {author} {\bibfnamefont {R.}~\bibnamefont {Golingo}}, \bibinfo {author} {\bibfnamefont {D.~D.}\ \bibnamefont {Hartog}}, \bibinfo {author} {\bibfnamefont {M.}~\bibnamefont {Hughes}}, \bibinfo {author} {\bibfnamefont {S.}~\bibnamefont {Knecht}}, \bibinfo {author} {\bibfnamefont {W.}~\bibnamefont {Lowrie}}, \bibinfo {author} {\bibfnamefont {V.}~\bibnamefont {Lukin}}, \bibinfo {author} {\bibfnamefont {B.}~\bibnamefont {Nelson}}, \bibinfo {author} {\bibfnamefont {R.}~\bibnamefont {Oberto}}, \bibinfo {author} {\bibfnamefont {J.}~\bibnamefont {Rohrbach}}, \bibinfo {author} {\bibfnamefont {M.}~\bibnamefont {Ross}}, \ and\ \bibinfo {author} {\bibfnamefont {G.}~\bibnamefont {Vogman}},\ }\bibfield  {title} {\enquote {\bibinfo {title} {The sheared-flow stabilized z-pinch},}\ }\href {\doibase 10.13182/FST12-A13407} {\bibfield  {journal} {\bibinfo  {journal} {Fusion
  Science and Technology}\ }\textbf {\bibinfo {volume} {61}},\ \bibinfo {pages} {119--124} (\bibinfo {year} {2012})},\ \Eprint {http://arxiv.org/abs/https://doi.org/10.13182/FST12-A13407} {https://doi.org/10.13182/FST12-A13407} \BibitemShut {NoStop}%
\bibitem [{\citenamefont {Golingo}\ and\ \citenamefont {Shumlak}(2003)}]{Golingo_RSI_2003}%
  \BibitemOpen
  \bibfield  {author} {\bibinfo {author} {\bibfnamefont {R.~P.}\ \bibnamefont {Golingo}}\ and\ \bibinfo {author} {\bibfnamefont {U.}~\bibnamefont {Shumlak}},\ }\bibfield  {title} {\enquote {\bibinfo {title} {Spatial deconvolution technique to obtain velocity profiles from chord integrated spectra},}\ }\href {\doibase 10.1063/1.1556956} {\bibfield  {journal} {\bibinfo  {journal} {Review of Scientific Instruments}\ }\textbf {\bibinfo {volume} {74}},\ \bibinfo {pages} {2332--2337} (\bibinfo {year} {2003})},\ \Eprint {http://arxiv.org/abs/https://doi.org/10.1063/1.1556956} {https://doi.org/10.1063/1.1556956} \BibitemShut {NoStop}%
\bibitem [{\citenamefont {Ross}\ and\ \citenamefont {Shumlak}(2016)}]{Ross2016RSI}%
  \BibitemOpen
  \bibfield  {author} {\bibinfo {author} {\bibfnamefont {M.~P.}\ \bibnamefont {Ross}}\ and\ \bibinfo {author} {\bibfnamefont {U.}~\bibnamefont {Shumlak}},\ }\bibfield  {title} {\enquote {\bibinfo {title} {{Digital holographic interferometry employing Fresnel transform reconstruction for the study of flow shear stabilized Z-pinch plasmas}},}\ }\href {\doibase 10.1063/1.4964387} {\bibfield  {journal} {\bibinfo  {journal} {Review of Scientific Instruments}\ }\textbf {\bibinfo {volume} {87}},\ \bibinfo {pages} {103502} (\bibinfo {year} {2016})},\ \Eprint {http://arxiv.org/abs/https://doi.org/10.1063/1.4964387} {https://doi.org/10.1063/1.4964387} \BibitemShut {NoStop}%
\bibitem [{\citenamefont {Zhang}\ \emph {et~al.}(2019)\citenamefont {Zhang}, \citenamefont {Shumlak}, \citenamefont {Nelson}, \citenamefont {Golingo}, \citenamefont {Weber}, \citenamefont {Stepanov}, \citenamefont {Claveau}, \citenamefont {Forbes}, \citenamefont {Draper}, \citenamefont {Mitrani}, \citenamefont {McLean}, \citenamefont {Tummel}, \citenamefont {Higginson},\ and\ \citenamefont {Cooper}}]{Zhang_2019_PRL}%
  \BibitemOpen
  \bibfield  {author} {\bibinfo {author} {\bibfnamefont {Y.}~\bibnamefont {Zhang}}, \bibinfo {author} {\bibfnamefont {U.}~\bibnamefont {Shumlak}}, \bibinfo {author} {\bibfnamefont {B.~A.}\ \bibnamefont {Nelson}}, \bibinfo {author} {\bibfnamefont {R.~P.}\ \bibnamefont {Golingo}}, \bibinfo {author} {\bibfnamefont {T.~R.}\ \bibnamefont {Weber}}, \bibinfo {author} {\bibfnamefont {A.~D.}\ \bibnamefont {Stepanov}}, \bibinfo {author} {\bibfnamefont {E.~L.}\ \bibnamefont {Claveau}}, \bibinfo {author} {\bibfnamefont {E.~G.}\ \bibnamefont {Forbes}}, \bibinfo {author} {\bibfnamefont {Z.~T.}\ \bibnamefont {Draper}}, \bibinfo {author} {\bibfnamefont {J.~M.}\ \bibnamefont {Mitrani}}, \bibinfo {author} {\bibfnamefont {H.~S.}\ \bibnamefont {McLean}}, \bibinfo {author} {\bibfnamefont {K.~K.}\ \bibnamefont {Tummel}}, \bibinfo {author} {\bibfnamefont {D.~P.}\ \bibnamefont {Higginson}}, \ and\ \bibinfo {author} {\bibfnamefont {C.~M.}\ \bibnamefont {Cooper}},\ }\bibfield  {title} {\enquote {\bibinfo {title} {Sustained neutron
  production from a sheared-flow stabilized $z$ pinch},}\ }\href {\doibase 10.1103/PhysRevLett.122.135001} {\bibfield  {journal} {\bibinfo  {journal} {Phys. Rev. Lett.}\ }\textbf {\bibinfo {volume} {122}},\ \bibinfo {pages} {135001} (\bibinfo {year} {2019})},\ \Eprint {http://arxiv.org/abs/https://doi.org/10.1103/PhysRevLett.122.135001} {https://doi.org/10.1103/PhysRevLett.122.135001} \BibitemShut {NoStop}%
\bibitem [{\citenamefont {Stangeby}(2000)}]{stangeby_2000}%
  \BibitemOpen
  \bibfield  {author} {\bibinfo {author} {\bibfnamefont {P.~C.}\ \bibnamefont {Stangeby}},\ }\href@noop {} {\emph {\bibinfo {title} {The plasma boundary of magnetic fusion devices}}}\ (\bibinfo  {publisher} {Institute of Physics Pub.},\ \bibinfo {year} {2000})\ \Eprint {http://arxiv.org/abs/https://doi.org/10.1201/9780367801489} {https://doi.org/10.1201/9780367801489} \BibitemShut {NoStop}%
\bibitem [{\citenamefont {Dasent}(1982)}]{dasent1982inorganic}%
  \BibitemOpen
  \bibfield  {author} {\bibinfo {author} {\bibfnamefont {W.}~\bibnamefont {Dasent}},\ }\href {https://books.google.com/books?id=sD85AAAAIAAJ} {\emph {\bibinfo {title} {Inorganic Energetics: An Introduction}}},\ Cambridge Texts in Chemistry and Biochemistry\ (\bibinfo  {publisher} {Cambridge University Press},\ \bibinfo {year} {1982})\BibitemShut {NoStop}%
\bibitem [{\citenamefont {Nemchinsky}(2014)}]{Nemchinsky_2014}%
  \BibitemOpen
  \bibfield  {author} {\bibinfo {author} {\bibfnamefont {V.}~\bibnamefont {Nemchinsky}},\ }\bibfield  {title} {\enquote {\bibinfo {title} {Erosion of thermionic cathodes in welding and plasma arc cutting systems},}\ }\href {\doibase 10.1109/TPS.2013.2287794} {\bibfield  {journal} {\bibinfo  {journal} {IEEE Transactions on Plasma Science}\ }\textbf {\bibinfo {volume} {42}},\ \bibinfo {pages} {199--215} (\bibinfo {year} {2014})},\ \Eprint {http://arxiv.org/abs/https://doi.org/10.1109/TPS.2013.2287794} {https://doi.org/10.1109/TPS.2013.2287794} \BibitemShut {NoStop}%
\end{thebibliography}%
\end{document}